\documentclass[prodmode]{acmsmall} % Aptara syntax

\usepackage[ruled]{algorithm2e}
\renewcommand{\algorithmcfname}{ALGORITHM}
\SetAlFnt{\small}
\SetAlCapFnt{\small}
\SetAlCapNameFnt{\small}
\SetAlCapHSkip{0pt}
\IncMargin{-\parindent}

\usepackage{graphicx}
\usepackage{amsfonts}
\usepackage{enumerate}
\usepackage{booktabs}
\usepackage{amsmath}
\usepackage{caption}
\usepackage{multirow}
\usepackage{subfigure}
\usepackage{float}
\usepackage{threeparttable}

\makeatletter
\def\runningfoot{}
\def\firstfoot{}
\makeatother

\begin{document}

\markboth{G. Wang et al.}{Ensemble Learning Based Classification
	Algorithm Recommendation}

% Title portion
\title{Ensemble Learning Based Classification Algorithm Recommendation}
\author{Guangtao Wang, Qinbao Song, Xiaoyan Zhu, and Jiao Liu
	\affil{Department of Computer Science and Technology, Xi'an Jiaotong University, China}}
% NOTE! Affiliations placed here should be for the institution where the
%       BULK of the research was done. If the author has gone to a new
%       institution, before publication, the (above) affiliation should NOT be changed.
%       The authors 'current' address may be given in the "Author's addresses:" block (below).
%       So for example, Mr. Abdelzaher, the bulk of the research was done at UIUC, and he is
%       currently affiliated with NASA.

\begin{abstract}
Selecting an appropriate classification algorithm for a given data set remains a challenging problem in data mining and machine learning. Existing algorithm recommendation models are typically trained with individual learners and rely on only one type of meta-feature, which may limit their ability to capture the diverse characteristics of classification problems. This paper proposes a multi-view ensemble meta-learning framework for classification algorithm recommendation. The framework constructs base recommendation models from different combinations of heterogeneous meta-feature groups and combines them through an accuracy- and diversity-aware ensemble strategy. The main focus of this work is empirical: we evaluate the proposed method on 1,090 benchmark classification problems derived from 84 public data sets, using 13 widely used candidate classification algorithms and five types of meta-features. The experimental results show that the proposed ensemble recommendation method consistently improves ranking loss, average precision, and top-ranked recommendation precision over individual recommendation models. These results suggest that combining complementary meta-feature views is an effective strategy for robust classification algorithm recommendation.
\end{abstract}

%\category{H.2.8}{Database Applications}{Data Mining}

\terms{Design, Algorithms, Performance}


%\acmformat{Gang Zhou, Yafeng Wu, Ting Yan, Tian He, Chengdu Huang, John A. Stankovic,
%and Tarek F. Abdelzaher, 2010. A multifrequency MAC specially
%designed for  wireless sensor network applications.}
% At a minimum you need to supply the author names, year and a title.
% IMPORTANT:
% Full first names whenever they are known, surname last, followed by a period.
% In the case of two authors, 'and' is placed between them.
% In the case of three or more authors, the serial comma is used, that is, all author names
% except the last one but including the penultimate author's name are followed by a comma,
% and then 'and' is placed before the final author's name.
% If only first and middle initials are known, then each initial
% is followed by a period and they are separated by a space.
% The remaining information (journal title, volume, article number, date, etc.) is 'auto-generated'.

%\begin{bottomstuff}
%This work is supported by the National Science Foundation, under
%grant CNS-0435060, grant CCR-0325197 and grant EN-CS-0329609.
%
%Author's addresses: G. Zhou, Computer Science Department,
%College of William and Mary; Y. Wu  {and} J. A. Stankovic,
%Computer Science Department, University of Virginia; T. Yan,
%Eaton Innovation Center; T. He, Computer Science Department,
%University of Minnesota; C. Huang, Google; T. F. Abdelzaher,
%(Current address) NASA Ames Research Center, Moffett Field, California 94035.
%\end{bottomstuff}

\maketitle

\section{Introduction}\label{sec:intro}

Classification is one of the most fundamental and widely used techniques in data mining and machine learning. Over the past decades, many classification algorithms have been developed, and new variants continue to emerge. However, both empirical studies \cite{brazdil2003ranking,Bensusan1998god,ali2006learning,smith2008cross,yang2006algorithm,song2012automatic,prudencio2011selecting,ali2018case,khan2020literature} and theoretical analyses \cite{wolpert2001supervised} show that no single classification algorithm is best suited to all classification problems. Instead, the most appropriate algorithm often depends on the characteristics of the given data set. As a result, users are frequently faced with the difficult task of selecting an effective algorithm from a large set of candidates. This motivates the need for automatic classification algorithm recommendation methods that can help users identify suitable algorithms for new problems.

Algorithm recommendation aims to model the relationship between the characteristics of classification problems and the algorithms that perform well on them. The learned relationship can then be used to recommend appropriate algorithms for a new classification problem. To facilitate the presentation, we introduce the following notation.
\begin{enumerate}[$\diamond$]
    \item $\mathbb{P}$ = \{$p_1$, $p_2$, $\cdots$, $p_N$\} denotes a collection of $N$ historical classification problems;
    \item $\mathbb{A}$ = \{$A_1$, $A_2$, $\cdots$, $A_k$\} denotes a collection of $k$ candidate classification algorithms;
    \item $F: \mathbb{P}\mapsto R^m$ denotes a function that extracts $m$ characteristics from each classification problem $p_i \in \mathbb{P}$ as the meta-features of $p_i$, where $R^m$ denotes the meta-feature space;
    \item $X = \{X_1, X_2, \cdots, X_{N}\} \subseteq R^m$ denotes the meta-features of the $N$ classification problems in $\mathbb{P}$ collected by $F$, where $X_i$ ($1\leq i\leq N$) is the meta-feature vector of $p_i$, i.e., $X_i = F(p_i)$;
    \item $Y = \{Y_1, Y_2, \cdots, Y_{N}\}$ denotes the meta-targets of the $N$ classification problems in $\mathbb{P}$, where $Y_i$ represents the algorithms in $\mathbb{A}$ that are appropriate for $p_i\in \mathbb{P}$.
\end{enumerate}

With this notation, \emph{algorithm recommendation} can be formally defined as a two-step process: (i) learning a function $\phi: X \mapsto Y$ that maps the characteristics of classification problems to appropriate classification algorithms; and (ii) for a new problem $p_{new}$, recommending algorithms according to $\phi(F(p_{new}))$.

The key challenge is therefore to construct an accurate function $\phi$, namely an algorithm recommendation model, that captures the relationship between the input meta-features $X$ and the output meta-targets $Y$. In algorithm recommendation, different representations of $Y$ lead to different ways of learning $\phi$. Common representations include single-label, algorithm-ranking, and multi-label meta-targets \cite{wang2014generic}. A single-label meta-target assumes that only the best-performing algorithm is appropriate for a given classification problem. An algorithm-ranking meta-target orders all candidate algorithms according to their performance. However, in practice, multiple classification algorithms are often statistically equivalent on the same problem under a given performance metric \cite{wang2014generic}. In such cases, neither the single-label nor the ranking-based representation is fully satisfactory. A multi-label meta-target provides a more natural representation because it allows multiple algorithms to be treated as appropriate for the same classification problem. Experimental results in \cite{wang2014generic} also show that recommendation models constructed from multi-label meta-data can achieve better performance. Therefore, this paper adopts the multi-label meta-target formulation and learns $\phi$ using multi-label learning methods.

A variety of supervised learning methods have been used to learn $\phi$ \cite{peng2002improved,Pfahringer00meta,kalousis2002algorithm,brazdil2003ranking,ali2006learning,prudencio2011selecting,song2012automatic,wang2014generic}, including single-label learning, \emph{k}-nearest-neighbor methods, and multi-label learning. Most early recommendation models were constructed with individual learners, while more recent work has begun to explore ensemble-based algorithm recommendation. In particular, \cite{zhu2021ensemble} proposed EML, an ensemble of ML-KNN models that leverages different meta-feature sets for classification algorithm recommendation. This paper follows the empirical motivation of using ensemble learning for algorithm recommendation, but studies a more general multi-view ensemble framework: base recommendation models are constructed from heterogeneous meta-feature combinations and are further selected through accuracy- and diversity-aware filtering before model combination.

Existing recommendation methods also typically learn $\phi$ from only one type of meta-feature. However, different types of meta-features characterize classification problems from different perspectives \cite{brazdil2003ranking,gama2000cascade,Engels98usinga,Bensusan1998god,peng2002improved,Pfahringer00meta,Bensusan2000casa,ho2002complexity,song2012automatic}. These meta-feature groups can be complementary: each may capture different aspects of a problem that are relevant to algorithm performance. By constructing recommendation models from different meta-feature combinations, we can obtain multiple complementary base models. Integrating these models provides an ensemble recommendation model that differs from existing single-feature and single-learner approaches. The experimental results reported in this paper demonstrate the effectiveness of the proposed ensemble learning-based recommendation method.

The main contributions of this paper are summarized as follows:
\begin{itemize}
    \item We formulate classification algorithm recommendation as a multi-label meta-learning problem, allowing multiple statistically competitive algorithms to be recommended for the same data set.
    \item We propose a multi-view ensemble recommendation framework that constructs base models from different combinations of heterogeneous meta-feature groups.
    \item We introduce an accuracy- and diversity-aware filtering strategy for selecting base recommendation models before ensemble combination. The theoretical discussion is used as motivation for this design rather than as a universal guarantee of ensemble superiority.
    \item We conduct an extensive empirical study on 1,090 classification problems, 13 candidate algorithms, and five types of meta-features, and show that the proposed ensemble strategy improves recommendation performance over individual recommendation models.
\end{itemize}

The rest of this paper is organized as follows. Section \ref{sec:relatedwork} discusses related work. Section \ref{sec:prestudy} introduces the preliminary study of ensemble learning-based algorithm recommendation. Section \ref{sec:ensembleAlgRec} presents the proposed ensemble learning-based recommendation method. Section \ref{sec:ExpStudy} reports the experimental study. Section \ref{sec:conclusion} concludes the paper.

\section{Related Work}\label{sec:relatedwork}

Algorithm recommendation has been formally studied as a meta-level
learning problem since the 1970s \cite{rice1975algorithm,ali2018case,khan2020literature}.
This idea has since become a common framework for algorithm selection
and meta-learning \cite{smith2008cross}. The input to this meta-level
learning problem is a set of characteristics of a classification problem
(i.e., ``meta-features''), and the output is the set of candidate algorithms
that are appropriate for the problem (i.e., the ``meta-target''). The meta-features
and meta-targets constitute the meta-data from which recommendation models
are constructed. From a data-mining perspective, the effectiveness of an
algorithm recommendation model depends mainly on the following two aspects.
\begin{enumerate}
    \item Meta-data preparation

    The recommendation model is induced from the meta-data. The
    quality of the meta-features and meta-target is therefore critical
    for model construction, following the well-known principle of
    ``Garbage In, Garbage Out'' in data mining \cite{lee1999cleansing}.
    Improving the quality of meta-data has consequently been a major
    direction in algorithm recommendation, and much prior work has focused
    on high-quality meta-data collection.

    \item Model construction

    The recommendation model is usually built with a given learning technique
    (e.g., classification) on the meta-data. Different learning
    techniques result in different recommendation models.
    In order to guarantee the generalization ability of the recommendation model,
    elaborate design of the learning schedule is another important research direction
    in algorithm recommendation.
\end{enumerate}

Many studies have investigated how to construct more effective
recommendation models
\cite{brazdil2003ranking,gama2000cascade,Engels98usinga,Bensusan1998god,peng2002improved,Pfahringer00meta,Bensusan2000casa,jain2000statistical,duin2004characterization,bernado2005domain,ho2002complexity,elizondo2009estimation,ho2000complexity,song2012automatic}.
These works mainly focus on the first aspect, including i)
extracting a set of high-quality meta-features to characterize a
classification problem and ii) developing an effective form of
meta-target to represent the appropriate algorithms on a
classification problem.

For meta-feature collection, different types of data set
characterization methods have been proposed from different
perspectives of a classification problem, including i) statistical and
information-theoretic methods, which extracts the statistic (e.g.,
mean value, standard deviation etc.) and information theory based
measures (e.g., entropy,  signal to noise ratio, etc.) as the
meta-features
\cite{brazdil2003ranking,gama2000cascade,Engels98usinga}; ii) model
structure based method, which first maps the classification problem
into a special data structure (e.g., decision tree) and then
extracts the properties of the structure as the meta-features
\cite{Bensusan1998god,peng2002improved}; iii) land-marking based
method, which characterizes a classification problem by the
performance metrics of a set of simple learners (also referred to as
land-marker) on the problem
\cite{Pfahringer00meta,Bensusan2000casa,jain2000statistical,duin2004characterization};
iv) problem complexity based method, which extracts a set of
measures reflecting the source of the difficulty to solve a
classification problem as the meta-features
\cite{ho2002complexity,elizondo2009estimation}; and v) structural
information based method, which uses structural information based
feature vectors to characterize the classification problems
\cite{song2012automatic}. All these meta-features have been employed
to construct algorithm recommendation models and given us some
useful guidelines for picking up appropriate algorithms.

Nevertheless, these recommendation models are usually constructed on
a single kind of meta-features by a single learner. As we know, an
ensemble learner combining a set of single learners in a specific
way (e.g., weighted or unweighted voting) is usually much more
accurate than the single learners
\cite{dietterichl2002ensemble,dietterich2000ensemble,dvzeroski2004combining}.
However, relatively little work has studied how to construct algorithm
recommendation models using ensemble learners.

For meta-target preparation, the expression form of the meta-target
has a great influence on the single learners used to build a
recommendation model. Two most widely used forms are single-label
\cite{ali2006meta,ali2006learning,kalousis2002algorithm,kalousis2004data}
and algorithm ranking
\cite{brazdil2003ranking,brazdil2000comparison,song2012automatic}.
The former assumes that there is a single optimal algorithm for a
given classification problem and forms a single-label meta-data.
Furthermore, the single-label learners are employed to build the
recommendation models. The latter ranks all candidate algorithms
according to their performance on a classification problem and
further gets a rank list of these algorithms as the meta-target.
Constrained by the ranking structure, the recommendation models on
algorithm-ranking-based meta-data are usually constructed by
\emph{k} nearest neighbor method or its variants. Recently, Wang et
al. \cite{wang2014generic} proposes a new and natural multi-label
form to describe the meta-target due to the fact that there would be
multiple algorithms that are appropriate for a given classification
problem in practice, and further constructs the recommendation model
by multi-label learning methods. The experimental results show that
the multi-label-based meta-target is more effective than the
single-label and ranking based ones. However, the multi-label learning
methods used in \cite{wang2014generic} are still based on individual
learners. This paper keeps the multi-label meta-target formulation but
constructs the recommendation model using multi-label ensemble learning.

Zhu et al. \cite{zhu2021ensemble} proposed EML, an ensemble of ML-KNN for classification algorithm recommendation. EML constructs a two-layer recommendation framework to leverage the diversity of different meta-feature sets and can automatically recommend different numbers of appropriate algorithms for different data sets. Their work is closely related to ours because both methods treat algorithm recommendation as a multi-label learning problem and exploit multiple meta-feature views. However, EML is built around ML-KNN as the underlying multi-label learner, whereas our framework constructs base recommendation models from different combinations of heterogeneous meta-feature groups and uses an accuracy- and diversity-aware filtering strategy before model combination. Thus, our method provides a more general ensemble construction strategy for multi-view algorithm recommendation.

Several studies have investigated ensemble learning for multi-label
classification
\cite{tsoumakas2007random,Gulisong2010triple,Lior2014ensemble,jesse2008multi,shi2011multi,min2014Review}.
These methods can be broadly grouped into two categories: data
transformation
\cite{tsoumakas2007random,Gulisong2010triple,Lior2014ensemble} and
ensemble adaptation \cite{shi2011multi,zhu2020automatic}. For the data transformation
methods, a multi-label problem is usually divided into multiple
single-label problems, then each base model of the ensemble is
trained on one of the single-label problems. The representative
method is RAndom k-labELsets (RAKEL) algorithm
\cite{tsoumakas2007random}, which transforms a multi-label learning
problem into an ensemble of a set of multi-class single-label
learning problems. The ensemble adaptation methods extend the
single-label ensemble learning method in order to handle the
multi-label problem directly. For example, Shi et al.
\cite{shi2011multi} proposed two multi-label-based criteria to
evaluate the accuracy and diversity of multi-label learning models
and then constructed an ensemble by optimizing these two criteria
with an Evolutionary Algorithm (EA). Since different base learners
will lead to different genetic representations and operations, and
not all the multi-label base learners can be optimized by EA, the
main weakness of their method is that it is mostly tailored for a
specific multi-label learners (e.g., BP-MLL \cite{zhang2006multi}
and ML-RBF \cite{zhang2009ml}), and thus lacks generality.

In this paper, we use a data transformation strategy to build
multi-label learning-based algorithm recommendation models because
it is simple to implement and is not tied to a specific learner. The
multi-label meta-data are transformed into multiple single-label
meta-data sets in both the meta-feature and meta-target spaces, and
the recommendation models constructed on these transformed data sets
serve as the base models of the ensemble.

\section{Preliminary Study}\label{sec:prestudy}

This section motivates the use of ensemble learning for algorithm
recommendation and introduces the accuracy- and diversity-based model
selection criteria used by the proposed framework. The discussion is
intended to guide the empirical design of the method rather than to
provide a universal theoretical guarantee.

\subsection{Rationality and Feasibility}\label{subsec:rationality}

Ensemble learning often improves generalization because it can partly
alleviate two common limitations of individual learners: the statistical
problem and the representational problem
\cite{dietterichl2002ensemble,dietterich2000ensemble}. Both limitations
also arise when algorithm recommendation models are constructed from
limited meta-data.

For example, most of the published studies in algorithm
recommendation usually employed only dozens of classification
problems in $\mathbb{P}$ to explore the function $\phi$
\cite{brazdil2003ranking,king1995statlog,song2012automatic,brazdil1994characterizing,ali2006meta,ali2006learning,prudencio2011selecting,peng2002improved,kalousis2004data,kalousis2002algorithm,brodley1993addressing,Pfahringer00meta}.
In some studies, comparing to the size of input $X$ (i.e., the
number of the classification problems in $\mathbb{P}$), its
dimension (i.e., the number of meta-features) is usually relatively
large. Such as, only 12 classification problems but up to 19
meta-features used in \cite{king1995statlog}, and 32 classification
problems and 8 meta-features used in \cite{prudencio2011selecting},
etc. With a limited number of classification problems in $\mathbb{P}$,
the greater the number of meta-features used, the more difficult to
find the true function $\phi$. This is identified as the statistical
problem encountered by the single learners. And this problem is
usually very significant and serious in the field of algorithm
recommendation. In order to overcome this issue and further get
recommendation model with better generalization ability, ensemble
learning will be a good choice.

Moreover, in practice, the performance of a classification algorithm
on a given classification problem is related to many factors (or
meta-features) of the problem, and different factors play different
roles \cite{wang2014generic}. This results in that the true function
$\phi$ might be quite complex. To approximate a complex function
$\phi$, the single learners usually might be limited by its
representational ability. The ensemble learner can enrich the
representational ability of the single learners by combining them in
a special way, and further relieve the representational problem.
Such as, the Fisher's linear discriminant algorithm only searches in
the linear space. However, an ensemble of multiple linear learners can
approach a non-linear function. Consequently, in the circumstances
without any prior knowledge of the form of $\phi$, ensemble learning
will be a sensible candidate.

The discussion above motivates ensemble learning as a practical design
choice for algorithm recommendation. Based on standard observations in
ensemble learning \cite{dietterichl2002ensemble,dietterich2000ensemble,hansen1990neural},
we use the following sufficient condition as guidance for selecting base
recommendation models.
%=============================================================
\begin{corollary}\label{coro:condition}
A standard sufficient condition for constructing an effective
ensemble learning model is that the base learning models are
individually accurate and diverse\footnote{In ensemble learning,
independence among base learners is often discussed in terms of
diversity.}.
\end{corollary}
%=============================================================

According to Corollary \ref{coro:condition}, a key design goal is to
select base recommendation models that are both reasonably accurate and
sufficiently diverse. In this paper,
we try to build different base recommendation models with respect to
different types of meta-features, and then assemble these base
recommendation models together to form the ensemble recommendation
model. The feasibility of this idea is supported by the following observations.
%================================================================
\begin{enumerate}
    \item Accurate base model construction

    \quad There have been many different recommendation models constructed
    using different types of meta-features and single
    learning methods \cite{brazdil2003ranking,gama2000cascade,Engels98usinga,Bensusan1998god,peng2002improved,Pfahringer00meta,Bensusan2000casa,jain2000statistical,duin2004characterization,ho2002complexity,elizondo2009estimation,song2012automatic}.
    And we can view these models as the base
    recommendation models of the ensemble. Ensemble learning model has one quite good
    property that it does not require all the base learning models
    to be highly accurate. That is, it is usually achieved by combining a
    set of weak base learning models \cite{dietterichl2002ensemble,dietterich2000ensemble}. Although there exist some differences among
    the existing recommendation models, all these models can effectively narrow down the choices of the
    candidate classification algorithms, and have reasonable
    recommendation accuracy. This provides evidence that it is reasonable to
    construct a set of accurate base recommendation models of
    ensemble.

%====================================================================
\begin{figure}[!h]
    \centering
    \includegraphics[width=0.65\textwidth]{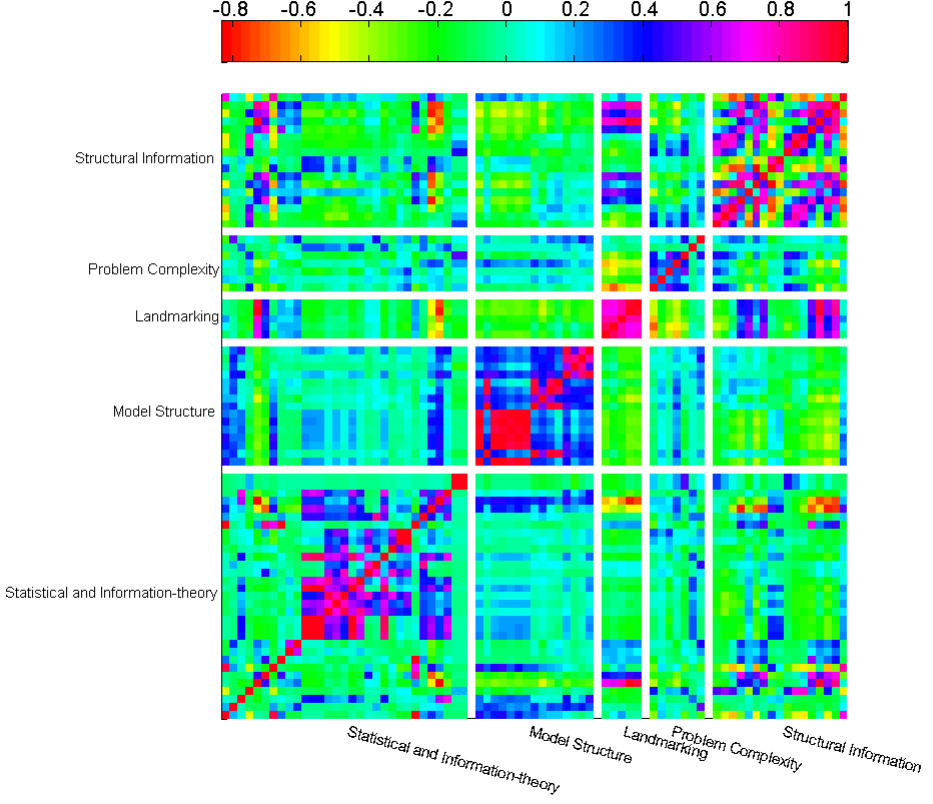}
    \caption{Correlation coefficients among different types of meta-features}\label{Fig:CorrelationAmongMetaFeatures}
\end{figure}
%====================================================================

    \item Diverse base model construction

    The literature has identified five different types of meta-features in the
    field of algorithm recommendation (See details in Appendix
\ref{appendix:meta-features}). These meta-features are extracted
    from different viewpoints of a classification problem independently. So it is
    reasonable to assume that, different types of meta-features are independent
    with each other. Fig. \ref{Fig:CorrelationAmongMetaFeatures} gives the correlation
    coefficients among the five kinds of meta-features extracted from
    1090 benchmark classification problems. From this figure, we can
    find that the correlation among different types of meta-features is
    usually quite low. This provides empirical evidence that different kinds of
    meta-features are independent of each other.
    Furthermore, it is more likely that different recommendation models constructed with
    these different types of meta-features will be independent/diverse.
\end{enumerate}
%================================================================

\subsection{Definitions of Accurate and Diverse Models}\label{subsec:accurateAndDiverseModel}

Let $M_{E}$ be an ensemble learning model constructed with $n$ base
learning models \{$M_1, M_2, \cdots, M_n$\}, and $pr_{M_i}$ ($1\leq
i\leq n$) be the probability of $M_{i}$ to make an error prediction
on a new coming instance. Then, for ensemble learning, an accurate
learning model can be defined as follows.

\begin{definition}\textbf{Accurate learning model}\label{def:accurateModel}.
A base learning model $M_i$ ($1\leq i\leq n$) is accurate if and
only if $pr_{M_i} < 1/2$.
\end{definition}

Definition \ref{def:accurateModel} tells us that a base learning
model is accurate if and only if its prediction error is less than
1/2. The rationality of this definition can be demonstrated as
follows.

\begin{enumerate}[$\diamond$]

\item First, we construct another ensemble learning model $\hat{M}_{E}$ over $n$ other
      base learning models \{$\hat{M}_1, \hat{M}_2, \cdots, \hat{M}_n$\}, where each base model
      $\hat{M}_i$ ($1\leq i\leq n$) has the identical probability to make an error
      prediction. And the probability is equal to the maximum value of \{$pr_{M_i}: 1\leq i\leq
      n$\}. i.e, $pr_{\hat{M}_i}$ = $\max\{pr_{M_1}, pr_{M_2}, \cdots, pr_{M_n}\}$ ($1\leq i\leq n$).

\item Then, by voting the predictions of the base learning models, we can get that
      $pr_{M_{E}} \leq pr_{\hat{M}_{E}}$ since $M_{E}$ is
      constructed over a set of base models with lower possibility to make an error
      prediction. Suppose that the base learning models of the ensemble learning model ($M_{E}$ or $\hat{M}_{E}$)
      are independent of each other, we can get a discrete random variable $X$ following
      binomial distribution ($n$, $pr_{\hat{M}_i}$), where $X$ indicates the
      number of models in \{$\hat{M}_1, \hat{M}_2, \cdots, \hat{M}_n$\}
      which make an error prediction. According to binomial distribution, we can get:
      \begin{equation}\label{eq:prerr}
        pr_{M_{E}} \leq pr_{\hat{M}_{E}} = Pr(X \geq \lceil n/2
        \rceil)
        = \sum\limits_{i=\lceil n/2
        \rceil}^{n}\binom{n}{i}(pr_{\hat{M}_i})^{i}(1-pr_{\hat{M}_i})^{n-i}.
      \end{equation}
      where $\lceil n/2 \rceil$ denotes the smallest integer greater than (or equal to) $n/2$.

\item Finally, according to Chernoff's inequality
      \cite{arratia1989tutorial}, for binomial distribution ($n, p$),
      \begin{equation}\label{eq:chernoff}
        pr_{M_{E}} \leq Pr(X \geq k) \leq \exp(-n\cdot D(\frac{k}{n}||p)), \mathrm{if\ and\ only\ if\ } p < \frac{k}{n} < 1.
      \end{equation}
      where $D(a||p)$ is the relative entropy between two Bernoulli distributions with
      parameters $a$ and $p$, and defined as $D(a||p) = a\cdot\log\frac{a}{p} + (1-a)\cdot\log\frac{1-a}{1-p}$.
      Corresponding to ensemble learning model $\hat{M}_{E}$, $p =
      pr_{\hat{M}_i}$, $k = \lceil n/2 \rceil$ and so $a = \frac{k}{n} \geq 1/2$.
      Therefore, if and only if $pr_{\hat{M}_i} < 1/2$, by voting the predictions of \{$\hat{M}_1, \hat{M}_2, \cdots, \hat{M}_n$\}
      as the prediction of $\hat{M}_{E}$, Eq. \ref{eq:chernoff} will be always true.
      And $pr_{M_{E}}$ will be bounded by $\exp(-n\cdot D(\frac{k}{n}||p))$.
      Moreover, in the case of $p = pr_{\hat{M}_i} < 1/2$, as $n$ increases, the value of $\exp(-n\cdot
      D(\frac{k}{n}||p))$ will approach 0 since $D(\frac{k}{n}||pr_{\hat{M}_i}) >0$. That is, the more base learning
      models used, the smaller the probability of $M_{E}$ to make an
      error prediction. This will be a very good property for ensemble
      learning. In a word, all these conclusions will be true under $pr_{\hat{M}_i} = \max\{pr_{M_1},
      pr_{M_2}, \cdots, pr_{M_n}\}$  < $1/2$. So in order to get an accurate
      ensemble learning model $M_{E}$ over \{$M_1, M_2, \cdots, M_n$\},
      $pr_{M_{i}}$ should be less than 1/2.
\end{enumerate}

Without the independence/diversity\footnote{In the field of ensemble
learning, the independence of base learners is generally called
diversity.} among the base models \{$M_1, M_2, \cdots, M_n$\}, the
random variable $X$ in Eq. \ref{eq:prerr} will not follow binomial
distribution and further the Eq. \ref{eq:chernoff} might be false.
This will result in that $pr_{M_{E}}$ might be non-convergent or
converging too slowly. This phenomenon has been recognized in
learner combination
\cite{cunningham2000diversity,lam2000classifier}. And the diverse
ensemble learner has a better potential to improve the accuracy than
non-diverse ensemble learner
\cite{peterson2005estimating,brown2005diversity,kuncheva2003measures}.
Therefore, there will be a notable question: ``How to define or
evaluate the independence/diversity among the base learning
models?''.

In the field of ensemble learning, it is usually difficult to
evaluate the diversity between different learning models directly.
Researchers usually resort to the prediction/classification results
of the learning models on a given test data. There have been several
metrics proposed based on the prediction results to assess the
diversity between different models
\cite{kuncheva2003measures,lee2013automatic,cunningham2000diversity,dietterich2000experimental}.
These metrics can guide us to identify the diverse base models.
Meanwhile, Kuncheva et al. \cite{kuncheva2003measures} have stated
that, in order to guarantee the improvement over the performance of
base models, there exists a minimum threshold value for each of
these diversity metrics to pick up the diverse base models for
ensemble learning.

Following these ideas, suppose that $M_i$ and $M_j$ ($1\leq i\neq\j
\leq n$) are two different base models, and $R_i$ and $R_j$ are the
prediction/classification results of $M_i$ and $M_j$ on a given
test data set $D$, we can give the definition of diverse learning
model for ensemble learning as follows.

\begin{definition}\textbf{Diverse learning model}\label{def:diverseModel}.
Two base models $M_i$ and $M_j$ are diverse with each other if and
only if $\psi(R_i, R_j) < \delta$, where $\psi$ is a function which
computes the diversity between $M_i$ and $M_j$ based on $R_i$ and
$R_j$, and $\delta$ is a given minimum threshold.
\end{definition}

In Definition \ref{def:diverseModel}, the computation of $\psi$
depends on the expression of the prediction results (e.g., $R_i$) of
a learning model. There are three general expressions of the
prediction results in the field of ensemble learning
\cite{kuncheva2003measures}.
\begin{enumerate}
    \item A numeric vector which records the predicted posterior probabilities
    of all class labels. e.g., for a classification problem with $k$ class
    labels, the prediction results are a vector with $k$ probability
    values \cite{Kagan1996Error,Kagan1999Linear}.
    \item Class label which directly indicates the predicted result \cite{dietterich2000ensemble,dietterich2000experimental,kohavi1996bias}.
    \item Correct/incorrect decision which records whether the predicted
    label is correct or not \cite{Ludmila2003Limit,Ho1998Random,Giacinto2000design}.
\end{enumerate}

For the numeric vector based expression, one of the assumptions is
that a learning model outputs independent estimates of the posterior
probabilities. However, this is usually not the case since all these
posterior probabilities sum up to a constant 1. Moreover, not all
the learning models can directly output the posterior probabilities
of the class labels. In the field of ensemble learning, the
researchers usually define the diversity function $\psi$ in terms of
either class label or correct/incorrect decision
\cite{kohavi1996bias,dietterich2000experimental,dietterich2000ensemble,Ludmila2003Limit,Ho1998Random,Giacinto2000design,kuncheva2003measures}.
In this paper, we propose a function, which makes full use of the
prediction results and considers both of the class label and
correct/incorrect decision, to pick up the diverse models for
ensemble in next section.

\subsection{Base Model Identification for Algorithm Recommendation}\label{subsec:modelSelection}

The definitions of accurate and diverse learning models in the previous
section are stated for single-label learning and provide useful guidance
for constructing ensembles over single-label learning problems. However, in this paper, we view algorithm
recommendation as a multi-label learning problem and attempt to
handle it by ensemble learning method.

Therefore, there will be a question: \emph{``How to identify the
accurate and diverse base models for algorithm recommendation with
respect to multi-label-based meta-data?''} And this question can be
answered by dividing it into the following ones.
\begin{enumerate}[$\diamond$]
    \item \emph{\textbf{Question 1}}: How to construct base recommendation models for ensemble on multi-label-based meta-data?
    \item \emph{\textbf{Question 2}}: How to identify an accurate recommendation model?
    \item \emph{\textbf{Question 3}}: How to identify a diverse recommendation model?
\end{enumerate}

\subsubsection{Answer to Question 1}
This paper employs the frequently-used multi-label ensemble learning
method, data transformation, to construct ensemble recommendation
model on the multi-label-based meta-data. That is, we first
transform the multi-label meta-data into multiple single-labeled
meta-data, and then build recommendation models on these
single-label meta-data as the base models of the ensemble.

Here, the process of multi-label meta-data transformation consists
of two steps: i) in meta-feature space, a number of different sets
of multi-label meta-data are generated with respect to different
combinations of the existing meta-features (See details in Section
\ref{subsec:DataCollection}); ii) in meta-target space, for each
multi-label meta-data generated in i), multiple different sets of
single-label meta-data are generated according to different labels
of the meta-target in a specific way (See details in Section
\ref{subsec:baseModel}).

\subsubsection{Answer to Question 2}
Once we achieve the base models, it is straightforward to identify
the accurate base models according to Definition
\ref{def:accurateModel}. That is, for a given base model, if its
classification error rate on test data is less than 1/2, it will be
accurate, otherwise not.

\subsubsection{Answer to Question 3}

In order to identify the diverse models, according to Definition
\ref{def:diverseModel}, two critical dimensions should be
considered: one is to find a function $\psi$ to evaluate the
diversity between two models, and the other is to set a proper
threshold $\delta$ to pick up the diverse models. However, the
existing researches usually just supply the function $\psi$ but no
effective approach to preassign the threshold $\delta$
\cite{kuncheva2003measures,lee2013automatic,cunningham2000diversity,dietterich2000experimental}.

In this paper, we present a statistical method that can not only
quantify the diversity between two models but also adaptively set
the threshold $\delta$. Different from the existing diversity
evaluation functions acting on the prediction results in terms of
either the class labels or the direct/indirect decisions, the
proposed diversity evaluation method concerning both of them, which
will make full use of prediction results.

Suppose there are two different learning models $M_1$ and $M_2$, and
a data set $D$ with $K$ class labels \{$C_1$, $C_2$, $\cdots$,
$C_K$\} ($K\geq2$), then we can construct a $K\times K$ contingency
table (see Table \ref{table:contingencyTable}) based on the class
labels and correct/incorrect decisions predicted by $M_1$ and $M_2$
on $D$.

%================================================================
\begin{table}[!h]
    \caption{The $K\times K$ contingency table}\label{table:contingencyTable}
    \centering
    \resizebox{0.55\textwidth}{!}{
    \begin{tabular}{c | c c c c |c}
    \hline
    Classified label & $C_1$ & $C_2$ & $\cdots$ & $C_K$ & Total \\
    \hline
    $C_1$ & $N_{1,1}$ & $N_{1,2}$ & $\cdots$ & $N_{1,K}$ & $N_{1,*}$\\
    $C_2$ & $N_{2,1}$ & $N_{2,2}$ & $\cdots$ & $N_{2,K}$ & $N_{2,*}$\\
    $\vdots$ & $\vdots$ & $\vdots$ & $\vdots$ & $\vdots$ & $\vdots$\\
    $C_K$ & $N_{K,1}$ & $N_{K,2}$ & $\cdots$ & $N_{K,K}$ & $N_{K,*}$\\
    \hline
    Total & $N_{*,1}$ & $N_{*,2}$ & $\cdots$ & $N_{*,K}$ & $N$\\
    \hline
    \end{tabular}
    }
\end{table}
%===========================================================================

In Table \ref{table:contingencyTable}, $N_{i,j}$ ($1\leq i, j\leq
K$) is the number of test instances incorrectly classified as $C_i$
and $C_j$ by $M_1$ and $M_2$, respectively. $N_{i,*} =
\sum\limits_{j = 1}^{K}N_{i,j}$ ($1\leq i\leq K$), $N_{*,j} =
\sum\limits_{i = 1}^{K}N_{i,j}$ ($1\leq j\leq K$) and $N =
\sum\limits_{i=1}^{K}N_{i,*} = \sum\limits_{j=1}^{K}N_{*,j} =
\sum\limits_{i=1}^{K}\sum\limits_{j=1}^{K}N_{i,j}$. $N$ is the total
number of instances of $D$ which are incorrectly predicted by either
$M_1$ or $M_2$.

With this $K\times K$ contingency table, we can get the diverse
measure $\kappa$ by Eq. \ref{eq:kappa}. The greater the value of
$|\kappa|$, the smaller the diversity between $M_1$ and $M_2$.

%==============================================================
\begin{equation}\label{eq:kappa}
    \kappa = \frac{\Theta_1 - \Theta_2}{1 - \Theta_{2}}.
\end{equation}
%==============================================================

where $\Theta_1 = \sum\limits_{i=1}^{K}\frac{N_{i,i}}{N}$ and
$\Theta_2 =
\sum\limits_{i=1}^{K}\frac{N_{i,*}}{N}\times\frac{N_{*,i}}{N}$.

According to the contingency table analysis, the joint frequency
distribution of $C_i$ and $C_j$ predicted by $M_1$ and $M_2$ is
$\frac{N_{i,j}}{N}$, and $\frac{N_{i,*}}{N}$ and $\frac{N_{*,j}}{N}$
correspond to the marginal frequency distributions of $C_i$ and
$C_j$. The rationality of $\kappa$ that can evaluate how strong
the independence between $M_1$ and $M_2$ is demonstrated as follows.

\begin{enumerate}
    \item Suppose that $M_1$ and $M_2$ are independent of each other,
    the expected joint distribution of $C_i$ and $C_j$ would be
    $\frac{N_{i,*}}{N}\times \frac{N_{*,j}}{N}$. In Eq. \ref{eq:kappa}, the numerator
    $\Theta_1 - \Theta_2$ can also be represented as
    $\sum\limits_{i=1}^{K}(\frac{N_{i,j}}{N} - \frac{N_{i,*}}{N}\times \frac{N_{*,j}}{N})$.
    Therefore, in the case that $M_1$ is independent of $M_2$,
    $\Theta_1 - \Theta_2$ will be quite close to 0 in practice.

    \item If the learner $M_1$ is positively related to $M_2$,
    an instance being predicted as $C_i$ ($1\leq i\leq K$) by $M_1$
    means that it is more likely that $M_2$ classifies the instance to $C_i$ as
    well. This will increase the value of $\frac{N_{i,i}}{N}$, i.e., the
    elements on the main diagonal of the contingency table.
    Otherwise, if $M_1$ is negatively related to or independent with $M_2$,
    the instance will be predicted as different classes by these two models.
    This will reduce the value of $\frac{N_{i,i}}{N}$. So we can
    get that the value of $N_{i,i}$ can reflect the dependence of two
    different learners. That is why we define the metric $\kappa$
    by the elements on the main diagonal of the contingency table.

   \item The denominator $1 - \Theta_{2}$ of $\kappa$ plays a role to limit
    the value of $\kappa$ into the range $[-1, 1]$. If the
    classification results on the test data are always identical,
    $\sum\limits_{i=1}^{K}{N_{i,i}} = N$, so $\Theta_1 = 1$ and
    $\kappa$ achieves its maximum value 1. If $M_1$ and $M_2$ are independent of each other,
    $\kappa$ will be 0 or quite near 0. And $\kappa < 0$/$> 0$ means $M_1$ is
    negative/positive related to $M_2$. The greater the value
    of $|\kappa|$, the stronger the dependence between $M_1$ and
    $M_2$.
\end{enumerate}

In practical application, the metric $\kappa$ is estimated via the
prediction results of the learning models on only a sample rather
than the whole population. This might also be the reason that there
needs a threshold $\delta$ in Definition \ref{def:diverseModel}.
Therefore, we need to further understand the statistical
significance of $\kappa$, including the statistical significance of
$\kappa\neq 0$ and its confidence interval. And the confidence
interval will be set as the threshold $\delta$ in Definition
\ref{def:diverseModel}.

For this purpose, we need to find a statistic for significant test
of $\kappa$. As we know, the distribution of the class labels
predicted by a learner on a $K-class$ classification problem would
follow either binomial ($K = 2$) or multi-nominal ($K > 2$)
distribution. Both of binomial, multi-nominal distributions are
derived from exponential family of distributions. Meanwhile,
inspiring by the idea that the independence between two variables,
which follow the well-known exponential distribution (i.e, normal
distribution), is usually statistically tested by a $t$-statistic,
we attempt to employ a $t$-statistic in Eq. \ref{eq:tstatistic} to
test the significance of $\kappa \neq = 0$, and further determine
whether two learning models $M_1$ and $M_2$ are independent with
each other or not.

%=================================================================
\begin{equation}\label{eq:tstatistic}
    t = \frac{\kappa}{\sqrt{\frac{1-\kappa^2}{N-2}}}
\end{equation}
%=================================================================

The $t$ statistic follows the Student's t-distribution with freedom
of degree $N-2$ under the null hypothesis that $M_1$ and $M_2$ are
independent of each other. If the $t$-statistic test accepts the
null hypothesis under given significance level $\alpha$, we can
conclude that the measure $\kappa$ has no significant difference
with 0, i.e., $M_1$ and $M_2$ are independent of each other.

According to Eq. \ref{eq:tstatistic}, its inverse can be calculated as
follow
\begin{equation}\label{eq:tinver}
    \kappa = \frac{t}{\sqrt{N - 2 + t^2}}
\end{equation}
where $t$ statistic follows student distribution with degree of
freedom $N-2$.

Let $t_{c}$ denote the critical value of student distribution with
degree of freedom $N-2$ under a given significance level $\alpha$
(e.g., $\alpha$ = 0.05), then we can get the confidence interval of
$\kappa = 0$ as $[-\frac{t_c}{\sqrt{N - 2 + t_c^2}},
\frac{t_c}{\sqrt{N - 2 + t_c^2}}]$. If $\kappa$ calculated between
two single learning models falls into this interval, we can conclude
that these two models are statistically independent of each other
under the given significance level $\alpha$. Based on this
confidence interval, we can set the minimum threshold $\delta =
\frac{t_c}{\sqrt{N - 2 + t_c^2}}$. And the diverse models can be
detected by comparing $\kappa$ with $\delta$ directly. That is, two
models $M_1$ and $M_2$ are independent of each other if and only
if $|\kappa| < \delta$.

\section{Ensemble Learning Based Algorithm Recommendation}\label{sec:ensembleAlgRec}

This section first shows the general view of the proposed ensemble
learning-based recommendation method, then describes the process of
model construction in detail.

\subsection{General View}\label{subsec:genView}

\begin{figure}[!h]
    \centering
    \includegraphics[width=0.75\textwidth]{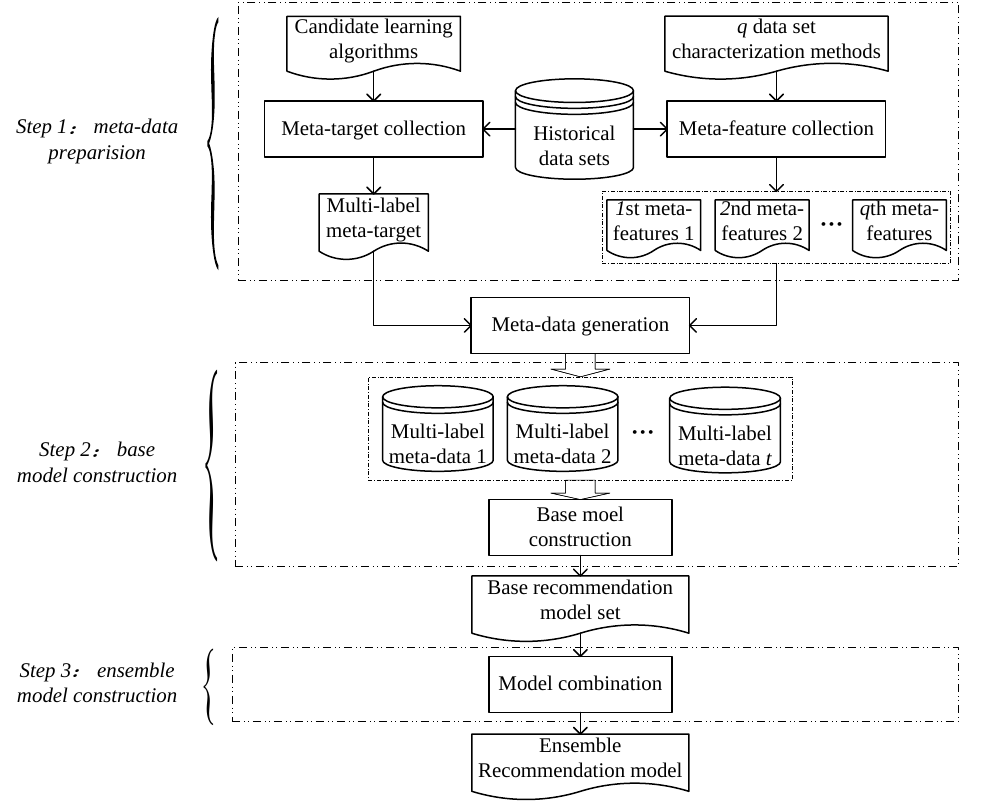}
    \caption{General view of ensemble recommendation model construction}\label{Fig:ModelConstructProcess}
\end{figure}

Firstly, different types of meta-features and the multi-labeled
meta-target are collected over a set of historical classification
problems; afterwards, by joining different combinations of these
meta-features and the multi-labeled meta-target together, different
sets of multi-label meta-data will be generated. Secondly, the
base recommendation models are constructed on each of the generated
multi-label meta-data. Thirdly, a multi-label ensemble learning
recommendation model will be achieved by combining these base
recommendation models together. Fig. \ref{Fig:ModelConstructProcess}
gives the general view of the proposed method which consists of
three steps: i) meta-data preparation, ii) base recommendation model
construction and iii) ensemble model construction.

%=================================================================
%=================================================================

\begin{enumerate}[1)]
    \item Meta-data preparation

    Meta-data is collected from a set of historical classification problems.
    i) For meta-feature collection, all the \emph{q} different kinds of data
    characterization methods are utilized on the historical classification problems to get \emph{q} groups of
    meta-features. ii) For meta-target collection, the
    appropriate algorithms of each historical classification problem are identified
    by statistically comparing all the candidate
    algorithms in terms of a given performance metric (such as, classification accuracy).
    And these appropriate algorithms form the multi-label-based meta-target.

    \quad  Different types of meta-features reflect the properties of a
    classification problem in different viewpoints. Combinations of these
    meta-features will give us a more comprehensive understanding
    of the problem. Inspired by this idea, this paper attempts to
    construct the base recommendation models with respect to
    different combinations of these meta-features in a specific way. Firstly, \emph{q} different
    sets of meta-features can be combined to generate $t = 2^q - 1$
    combinations (See the generation process in Section \ref{subsec:DataCollection}). Then, by merging these $t$ combinations and
    the multi-labeled meta-target together, $t$ different sets of multi-label meta-data can be
    generated.

    \item Base recommendation model construction

    For each multi-labeled meta data, firstly, the data transformation
    method is performed to transform the multi-labeled
    meta-data into multiple single-label meta-data, and then the
    base recommendation models will be generated from these
    single-label meta-data. The details of this process will be described in Section
    \ref{subsec:baseModel}.

    \item Ensemble recommendation model construction

    Once achieving the base recommendation models, the accurate and
    diverse base models are identified for constructing an ensemble
    recommendation model according to Section \ref{subsec:modelSelection}.
    For a new classification problem, the recommendations of
    these identified base models are combined in a specific way to form the
    recommended algorithms for the new problem. The detailed
    process of ensemble model construction will be introduced in
    Section \ref{subsec:ensembleModel}.
\end{enumerate}

\subsection{Meta-data Preparation}\label{subsec:DataCollection}

Let $\mathbb{A} = \{A_i: i = 1, 2, \cdots, k\}$ be a set of $k$
candidate classification algorithms, $\mathbb{P} = \{p_i: i = 1, 2,
\cdots, n\}$ be a set of $n$ historical classification problems, and
$F = \{F_i: i = 1, 2, \cdots, q\}$ be $q$ different data set
characterization functions used for meta-feature extraction.
$X_{i}^{j} = F_{j}(p_i)$ denotes the meta-features extracted by
$F_j$ ($1\leq j \leq q$) on the classification problem $p_i$ ($1\leq
i\leq n$).

Suppose that $D = \{(X_i, Y_i): i = 1, 2, \cdots, n\}$ denotes a set
of multi-label meta-instances, where $X_i$ is the meta-features
extracted from the classification problem $p_i$ by the function(s)
in $F$, that is, $X_i$ $\in$ $nchoosek(\{X_{i}^{1}$, $X_{i}^{2}$,
$\cdots$, $X_{i}^{q}\})$ and here $nchoosek(Z)$ outputs all the
possible combinations of the elements in $Z$. For example, let $Z =
\{a,b,c\}$, then $nchoosek(Z)$ = \{\{$a$\}, \{$b$\}, \{$c$\},
\{$a,b$\}, \{$a,c$\}, \{$b,c$\}, \{$a,b,c$\}\}; and $Y_i =
\{Y_{i,j}: j = 1, 2, \cdots, k\}$ represents the multi-label-based
meta-target on $p_i$ and $Y_{i,j} =$ 1 or 0 indicates the algorithm
$A_j$ is appropriate or inappropriate on $p_i$.

It is noted that, there are many methods to generate different
combinations of meta-features from the given $q$ kinds of
meta-features. The combination function $nchoosek()$ is chosen is
because it can not only generate multiple different
sets of meta-data for base model construction, but also help to find
whether the combinations of different types of meta-features are
better, and further discover the salient meta-features for algorithm
recommendation.

With the combination function $nchoosek()$ on $q$ different kinds of
meta-features, we can generate $t = 2^q - 1$ different sets of
meta-features and furthermore $t$ sets of multi-label meta-data.

\subsection{Base Recommendation Model Construction}\label{subsec:baseModel}

For each one of the $t$ sets of multi-label meta-data, the process
to construct the base recommendation models is identical. This
section will illustrate this process by taking one given
multi-label meta-data $D$ as an example.

%==================================================================
\begin{figure}[!h]
    \centering
    \includegraphics[width=0.75\textwidth]{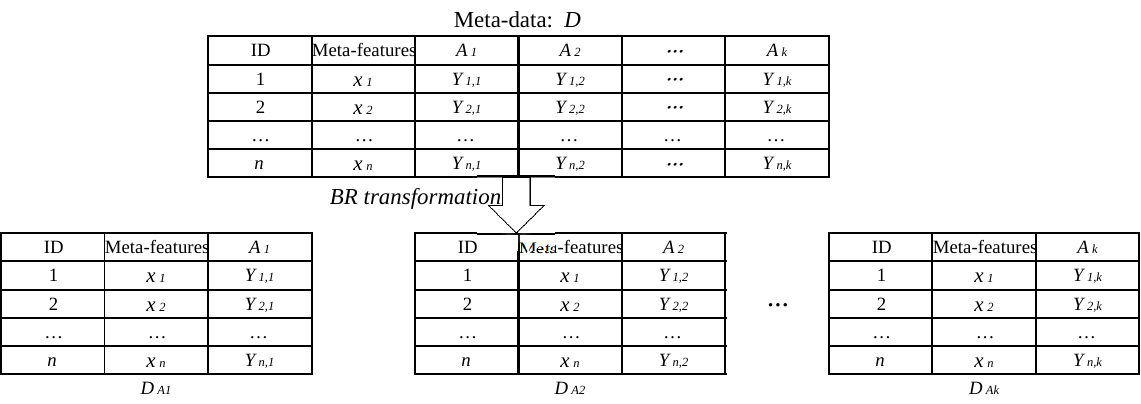}
    \caption{Data sets transformed by BR method}\label{Fig:dataTransformation}
\end{figure}
%==================================================================

At first, the multi-label meta-data $D$ is transformed into
multiple single-label data sets in meta-target space. The popular
data transformation method, Binary Relevance (BR)
\cite{tsoumakas2010mining}, is performed. The BR method transforms
$D$ into $k$ data sets $D_{A_j}, j= 1,\cdots,k$ which contain all
the instances of $D$ labeled by whether algorithm $A_j$ is
appropriate or not. This means that for each instance $(X_i, Y_i)$ of
$D$, the corresponding instance of $D_{A_j}$ is $(X_i, Y_{i,j})$.
Fig. \ref{Fig:dataTransformation} shows the $k$ single-label data
sets produced by BR on the multi-label meta-data $D$.

Afterwards, $k$ recommendation models will be learned on these $k$
binary learning data sets by a specific classification algorithm
(e.g, Decision Tree). These $k$ models $\{M_1, M_2, \cdots, M_k\}$
constitute the base recommendation models on $D$. Here, $M_i$ is a
learning model which can output the probability $pr_{i}$
($1\leq i\leq k$) of the candidate algorithm $A_i$ being
appropriate.

\subsection{Combination of Base Recommendation Models}\label{subsec:ensembleModel}

By combining the $q$ different types of meta-features, $t = 2^q - 1$
different sets of multi-label meta-data are achieved. After
constructing base recommendation models on each one of the $t$
multi-label meta-data, we will get a learning model matrix
$M_{t\times k}$, where $M_{i,j} (1\leq i\leq t, 1\leq j \leq k)$
denotes the model learned on the $j$th data set transformed from
$i$th multi-label meta-data by BR method.

%==============================================================
\begin{equation}
    M_{t\times k} =\left[
  \begin{array}{cccc}
    M_{1,1} & M_{1,2} & \cdots & M_{1,k} \\
    M_{2,1} & M_{2,2} & \cdots & M_{2,k} \\
    \vdots & \vdots &  \vdots & \vdots \\
    M_{t,1} & M_{t,2} & \cdots & M_{t,k} \\
  \end{array}
    \right]
\end{equation}
%==============================================================

Let $M_{\cdot,j} (1\leq j \leq k)$ denote the $j$th column of the
matrix $M_{t\times k}$. According to the process of BR
transformation method, $M_{\cdot,j}$ consists of all the $t$
learning models with respect to the candidate algorithm $A_j$. Thus,
it is possible for us to construct the ensemble learning model
improve the prediction of $A_j$ based on the $t$ models in
$M_{\cdot,j}$.

According to Corollary \ref{coro:condition}, a useful practical strategy is to
construct the ensemble from base models that are both accurate and diverse.
Thus, ensemble model construction can be viewed as a process of selecting
accurate and sufficiently diverse models from $M_{\cdot,j}$.

%================================================================
\IncMargin{0.5em} \RestyleAlgo{ruled}\LinesNumbered
\begin{algorithm}[!ht]
  \SetAlgoVlined
  \caption{ModelFilter()}\label{algo:modelFilter}
  %=========================================================
  \SetKwInOut{Input}{Inputs}
  \SetKwInOut{Output}{Output}
  \SetKwComment{tcc}{/*}{*/}
  \SetKwComment{rcc}{//}{}
  %=========================================================
 \Input{
    $M_{\cdot,j} = \{M_{1,j}, M_{2,j}, \cdots, M_{t,j}\}$\;
    \hspace{1.3cm} $Accs = \{acc_1, acc_2, \cdots, acc_t\}$\;
    \hspace{1.3cm} $Outs = \{out_1, out_2, \cdots, out_t\}$;
 }
%=========================================================
 \BlankLine
 \Output{ $FlagList = \{f_1, f_2, \cdots, f_t\}$;}
%=========================================================
 \BlankLine
 Initialization: $\{f_i = 1: (1\leq i\leq t)\}$\;
 \rcc{Part 1: Accuracy based filter}
 \For{$i = 1$ \KwTo $t$} {
    \If{$acc_i < 0.5$}{
        $f_i = 0$;
    }
 }
 \rcc{Part 2: Diversity based filter}
 $IX$ = sort($Accs$, ``descend'')\rcc{$IX$ is the indices of the models sorted by $Accs$ in a descending order}
 \For{$i = 1$ \KwTo $t-1$} {
    $Idi$ = $IX_i$;\rcc{The $i$th element of $IX$}
    \If{$f_{Idi} == 1$}{
        \For{$j = (i+1)$ \KwTo $t$} {
            $Idj$ = $IX_j$\rcc{The $j$th element of $IX$}
            \If{$f_{Idj} == 1$}{
                $\kappa$ = DiversityComp($Out_{Idi}$, $Out_{Idj}$)\rcc{According to Eq. \ref{eq:kappa} of $\kappa$}
                \If{$\kappa \geq \delta$}{
                    $f_{Idj} = 0$;\rcc{$\delta$ is set according to Eq. \ref{eq:tinver}}
                }
            }
        }
    }
 }
\end{algorithm}
\DecMargin{0.5em} \LinesNumbered
%================================================================

Algorithm \ref{algo:modelFilter} gives the filtering method on the
$t$ models in $M_{\cdot,j}$. In this algorithm, besides
$M_{\cdot,j}$, there are two other input variables $Accs$ and $Outs$
with respect to the models in $M_{\cdot,j}$, where $acc_i \in Accs$
($1\leq i \leq t$) denotes the classification accuracy of $M_{i,j}$
on a validation data\footnote{The validation data is drawn from the
original meta-data and never used for model construction.}, and
$out_i \in Outs$ represents the outputs (i.e., predicted labels and
correct/incorrect decisions) of $M_{i,j}$ on the validation data.
$FlagList$ records the filtering results, where $f_i \in \{0,1\}$
and $f_i = 0/1$ means that $M_{i,j}$ is filtered out/reserved for
ensemble model construction.

Algorithm \ref{algo:modelFilter} consists of two filters: i)
accuracy based filter (lines 2-3) and ii) diversity based filter
(lines 5-14). The first filter is used to find out the accurate
models. In this filter, the models whose classification accuracy
being smaller than 1/2 are filtered out (i.e., set the corresponding
flags in $FlagList$ as 0) according to Definition
\ref{def:accurateModel} of accurate model. The second filter aims at
finding out the diverse models. In this filter, if the $\kappa$
between two specified models is greater than the predefined
threshold $\delta$ according to Definition \ref{def:diverseModel} of
diverse model, the model with lower classification accuracy will be
filtered out. Where the threshold $\delta$ is set according to Eq.
\ref{eq:tinver}.

With the help of Algorithm \ref{algo:modelFilter}, we can get a flag
matrix $G_{t\times k}$, where the column $f_{\cdot,j} (1\leq j\leq
k)$ is achieved by applying Algorithm \ref{algo:modelFilter} on the
models $M_{\cdot, j}$ of the $j$th column in the model matrix
$M_{t\times k}$, $f_{i,j} \in \{0,1\}$ ($1\leq i\leq t$), and
$f_{i,j} = 1$ means the model $M_{i,j}$ is chosen for ensemble model
construction.
%==============================================================
\begin{equation}
    G_{t\times k} =\left[
  \begin{array}{cccc}
    f_{1,1} & f_{1,2} & \cdots & f_{1,k} \\
    f_{2,1} & f_{2,2} & \cdots & f_{2,k} \\
    \vdots & \vdots &  \vdots & \vdots \\
    f_{t,1} & f_{t,2} & \cdots & f_{t,k} \\
  \end{array}
    \right]
\end{equation}
%==============================================================

Meanwhile, let $M_{i,\cdot} (1\leq i\leq t)$ be the $i$th row of the
model matrix $M_{t,k}$, which consists of the models learned from
the $i$th multi-label meta-data. For a new coming classification
problem $p_{new}$, each model $M_{i,j} \in M_{i,\cdot}$ ($1\leq
j\leq k$) predicts the probability $pr_{i,j}$ to recommend the
algorithm $A_j$ to $p_{new}$. So we can further get a matrix
$P_{t\times k}$ of probabilities predicted by the model matrix
$M_{t\times k}$ on $p_{new}$ as follows.

%==============================================================
\begin{equation}
    P_{t\times k} =\left[
  \begin{array}{cccc}
    pr_{1,1} & pr_{1,2} & \cdots & pr_{1,k} \\
    pr_{2,1} & pr_{2,2} & \cdots & pr_{2,k} \\
    \vdots & \vdots &  \vdots & \vdots \\
    pr_{t,1} & pr_{t,2} & \cdots & pr_{t,k} \\
  \end{array}
    \right]
\end{equation}
%==============================================================

With the predicted probability matrix $P_{t\times k}$ and the
flag matrix $G_{t\times k}$, the ensemble recommendation model will
estimate the probability of the candidate algorithm $A_j
(1\leq j\leq k)$ that are appropriate on $p_{new}$ by the Eq.
\ref{eq:ensemPro}.

%===============================================================
\begin{equation}\label{eq:ensemPro}
    pr_{en}(A_j) = \frac{\sum\limits_{i=1}^{t}pr_{i,j}\cdot f_{i,j}}{\sum\limits_{i=1}^{t}f_{i,j}}
\end{equation}
%===============================================================

In Eq. \ref{eq:ensemPro}, the denominator
$\sum\limits_{i=1}^{t}f_{i,j}$ denotes the number of models chosen
by Algorithm \ref{algo:modelFilter} on $M_{\cdot,j}$, i.e., the
number of based models used for ensemble learning with respect to
algorithm $A_j$. The molecule of Eq. \ref{eq:ensemPro} can be viewed
as a kind of weighted voting ensemble, where $pr_{i,j}$ denotes the
weight of model $M_{i,j}$ and it works if and only if $f_{i,j} = 1$,
i.e., $M_{i,j}$ is picked up as a base model for ensemble learning.

Afterwards, the ensemble recommendation model can rank the $k$
candidate algorithms according to the estimated probabilitys
$\{pr_{en}(A_1), pr_{en}(A_2), \cdots, pr_{en}(A_k)\}$. Meanwhile,
the algorithm $A_j$ will be recommended as an appropriate one for
$p_{new}$ if and only if $pr_{en}(A_j)$ is greater than a specific
threshold (e.g., 1/2).

%================================================================
\begin{table}[!h]
    \caption{Example for ranking strategy}\label{table:RankExample}
    \centering
    \resizebox{0.65\textwidth}{!}{
    \begin{tabular}{l c c c c c}
    \hline
    Algorithm & $A_1$ & $A_2$ & $A_3$ & $A_4$ & $A_5$ \\
    \hline
    Probability value & 0.7 & 0.4 & 0.8 & 0.5 & 0.7 \\
    Ranks & 2 & 5 & 1 & 4 & 3 \\
    Ranks (considering ties) &2.5 & 5 & 1 & 4 & 2.5\\
    \hline
    \end{tabular}
    }
\end{table}
%===========================================================================

Meanwhile, these $k$ probabilitys can be further used to learn
a ranking list of the candidate algorithms in $\mathbb{A}$ on
$p_{new}$. The algorithm with the highest probability will be ranked
first, the algorithm with the second highest probability will be
ranked second, and so on. In case of ties, average ranks are
assigned. This strategy can be illustrated by the example in Table
\ref{table:RankExample} in which we give the estimated probability
values of five candidate algorithms and corresponding ranks.

\section{Experimental Study}\label{sec:ExpStudy}
This section presents the empirical evaluation of the proposed ensemble
learning-based algorithm recommendation method. Because the main goal of
this work is to demonstrate the effectiveness of combining heterogeneous
meta-feature views, the experiments are organized around the following
research questions:
\begin{enumerate}
    \item \textbf{RQ1}: Does the proposed ensemble recommendation model outperform individual recommendation models built from different meta-feature combinations?
    \item \textbf{RQ2}: Does combining heterogeneous meta-feature views improve recommendation performance compared with relying on a single view?
    \item \textbf{RQ3}: Does the proposed accuracy- and diversity-based filtering strategy improve the ensemble over simpler combination strategies?
    \item \textbf{RQ4}: Which types or combinations of meta-features contribute most to classification algorithm recommendation?
\end{enumerate}
We first describe the experimental setup and then analyze the results in terms of these questions.

%===========================================================================
\begin{table*}[!h]
    \caption{Description of the 84 classification problems}\label{table:dataset}
    \centering
    \resizebox{0.85\textwidth}{!}{
    \begin{tabular}{l l l l l | l l l l l}
    \hline
    ID  &  Name & \# Attributes & \# Instances  & \# Classes & ID  &  Name & \# Attributes & \# Instances  & \# Classes \\
    \hline
    1 & anneal & 38 & 898 & 6 & 43 & liver-disorders & 6 & 345 & 2\\
2 & anneal.ORIG & 38 & 898 & 6 & 44 & lung-cancer & 56 & 32 & 3\\
3 & arrhythmia & 279 & 452 & 16 & 45 & lymph & 18 & 148 & 4\\
4 & audiology & 69 & 226 & 24 & 46 & mfeat-fourier & 76 & 2000 & 10\\
5 & australian & 14 & 690 & 2 & 47 & mfeat-karhunen & 64 & 2000 & 10\\
6 & autos & 25 & 205 & 7 & 48 & mfeat-morphological & 6 & 2000 & 10\\
7 & balance-scale & 4 & 625 & 3 & 49 & mfeat-zernike & 47 & 2000 & 10\\
8 & breast-cancer & 9 & 286 & 2 & 50 & molecular-biology\_promoters & 58 & 106 & 2\\
9 & breast-w & 9 & 699 & 2 & 51 & monks-problems-1 & 6 & 556 & 2\\
10 & car & 6 & 1728 & 4 & 52 & monks-problems-2 & 6 & 601 & 2\\
11 & cleve & 11 & 303 & 2 & 53 & monks-problems-3 & 6 & 554 & 2\\
12 & cmc & 9 & 1473 & 3 & 54 & mushroom & 22 & 8124 & 2\\
13 & colic & 22 & 368 & 2 & 55 & nursery & 8 & 12960 & 5\\
14 & connect-4 & 42 & 13512 & 3 & 56 & optdigits & 64 & 5620 & 10\\
15 & credit-a & 15 & 690 & 2 & 57 & page-blocks & 10 & 5473 & 5\\
16 & credit-g & 20 & 1000 & 2 & 58 & pendigits & 16 & 10992 & 10\\
17 & crx & 15 & 690 & 2 & 59 & pima & 6 & 768 & 2\\
18 & cylinder-bands & 39 & 540 & 2 & 60 & postoperative-patient-data & 8 & 90 & 3\\
19 & dermatology & 34 & 366 & 6 & 61 & primary-tumor & 17 & 339 & 22\\
20 & diabetes & 8 & 768 & 2 & 62 & segment & 19 & 2310 & 7\\
21 & ecoli & 7 & 336 & 8 & 63 & shuttle-landing-control & 6 & 15 & 2\\
22 & flags & 29 & 194 & 8 & 64 & sick & 29 & 3772 & 2\\
23 & german & 15 & 1000 & 2 & 65 & solar-flare\_1 & 12 & 323 & 2\\
24 & glass & 9 & 214 & 7 & 66 & solar-flare\_2 & 12 & 1066 & 3\\
25 & haberman & 3 & 306 & 2 & 67 & sonar & 60 & 208 & 2\\
26 & hayes-roth & 4 & 132 & 3 & 68 & soybean & 35 & 683 & 19\\
27 & heart-c & 13 & 303 & 5 & 69 & spambase & 57 & 4601 & 2\\
28 & heart-h & 13 & 294 & 5 & 70 & spect & 22 & 267 & 2\\
29 & heart-statlog & 13 & 270 & 2 & 71 & spectrometer & 102 & 531 & 48\\
30 & hepatitis & 19 & 155 & 2 & 72 & splice & 61 & 3190 & 3\\
31 & horse-colic.ORIG & 21 & 368 & 2 & 73 & sponge & 45 & 76 & 3\\
32 & hypo & 23 & 3163 & 2 & 74 & tae & 5 & 151 & 3\\
33 & hypothyroid & 29 & 3772 & 4 & 75 & tic-tac-toe & 9 & 958 & 2\\
34 & ionosphere & 34 & 351 & 2 & 76 & trains & 32 & 10 & 2\\
35 & iris & 4 & 150 & 3 & 77 & transfusion & 3 & 748 & 2\\
36 & kdd\_JapaneseVowels\_1 & 14 & 5687 & 9 & 78 & vehicle & 18 & 846 & 4\\
37 & kdd\_JapaneseVowels\_2 & 13 & 4274 & 9 & 79 & vote & 16 & 435 & 2\\
38 & kdd\_synthetic\_control & 61 & 600 & 6 & 80 & vowel & 13 & 990 & 11\\
39 & kr-vs-kp & 36 & 3196 & 2 & 81 & waveform-5000 & 40 & 5000 & 3\\
40 & labor & 16 & 57 & 2 & 82 & wine & 13 & 178 & 3\\
41 & led7 & 7 & 3200 & 10 & 83 & yeast & 7 & 1484 & 10\\
42 & letter & 16 & 20000 & 26 & 84 & zoo & 17 & 101 & 7\\
    \hline
    \end{tabular}
    }
\end{table*}
%===========================================================================
\subsection{Experimental Setup}

To evaluate the effectiveness and practical applicability of the proposed
ensemble learning-based algorithm recommendation method, we use the
following experimental setup.

\subsubsection{Benchmark Classification Problem}

We use 84 widely used public classification problems from the UCI
repository\footnote{http://archive.ics.uci.edu/ml/datasets.html.}
in the experiments. Table \ref{table:dataset} shows the
statistical summary of these problems in terms of the number of
attributes, the number of instances and the number of classes.

Moreover, in order to guarantee the reliability and soundness of the
conclusion, more classification problems should be employed. Thus,
with the help of the problem generation method, Datasetoids, which
was proposed in \cite{soares2009uci} and aimed to obtain a large
number of classification problems for algorithm recommendation
\cite{prudencio2011selecting,prudencio2011uncertainty,halabi2012identifying,prudencio2011combining},
we extend the 84 publicly UCI classification problems into 1090 (84
data sets and 1006 datasetoids) different classification problems.
The Datasetoids method generates the new
classification problems by exchanging role of each nominal attribute
with that of a target concept, i.e., viewing the nominal attribute
as the new target concept.

\subsubsection{Meta-feature Collection}

Five different types of meta-features are extracted from the 1090
classification problems. They are i) the statistical and
information-theory based, ii) the model structure based, iii) the
landmarking based, iv) the problem complexity based and v) the
structural information based. See Appendix
\ref{appendix:meta-features} for the details.

\subsubsection{Meta-target Collection}

Meta-target tells us the appropriate candidate algorithms for each
of the 1090 classification problems. Next, we introduce the
candidate classification algorithms in the experiments and how to
identify the appropriate algorithms for a given classification
problem.

\begin{enumerate}
    \item \emph{Candidate classification algorithms}

    To improve the generality of the experimental
    results, 13 different types of classification algorithms are selected
    as candidates.

    \quad These algorithms include i) the probability based algorithm Bayes Network;
    ii) tree based algorithms C4.5, RandomTree and RandomForest; iii)
    rule-based algorithms PART, Ripper and NNge; iv) Gaussian function
    based algorithm RBFNetwork, and v) support vector machine based algorithm SMO.

    \quad Besides the above nine single classification algorithms, we also include two types of well-known
    ensemble classification algorithms: Boosting and Bagging. They are applied with the base
    classifiers Naive Bayes (NB) and C4.5, respectively.

    \item \emph{Appropriate algorithm identification}

    The multi-label-based meta-target indicating the appropriate algorithms for a classification
    problem $D$ can be expressed in terms of a binary value vector $B_{D} = <b_1, b_2,
    \cdots, b_{13}>$, where $b_i = 1$ means that the corresponding candidate
    algorithm $A_i$ ($1\leq i\leq 13$) is appropriate. The appropriate algorithms are
    identified by their performance metrics (i.e., classification accuracy) on $D$ as follows.

    \begin{enumerate}
        \item {Process of classification accuracy estimation on $D$}

        In order to get a stable estimation of classification accuracy of
        the candidate algorithms on $D$, $5\times10$-fold stratified cross-validation is performed as
        the following steps. i) The problem $D$ is randomly split into ten mutually
        exclusive subsets $D_1, D_2, \cdots, D_{10}$ of equal size, and
        $D = \bigcup_{j=1}^{10} D_j$. ii) $D - D_j$ and $D_j$ ($j \in
        \{1,2,\cdots,10\}$) are used as the training and test sets, respectively.
        Each algorithm $A_i$ ($1\leq i \leq 13$)
        is trained on $D - D_j$, and its classification
        accuracy is estimated on $D_j$. iii) Repeat i) and ii) five times
        on $D$ whose instances are randomly re-ordered. Afterwards,
        for each candidate algorithm $A_i$ ($1\leq i \leq 13$),
        we will get a vector $Acc_i = <acc_{i,1}, acc_{i,2},
        \cdots, acc_{i,50}>$ with 50 classification accuracies.

        \item{Binary-valued based meta-target $B_{D}$ identification}

        In order to identify the truly appropriate algorithms from
        13 candidate algorithms according to the
        collected performance sets $\{Acc_1$, $Acc_2$, $\cdots$, $Acc_{13}\}$ on $D$,
        the statistical algorithm selection is a reasonable and commonly-used approach
        \cite{pizarro2002multiple}.

        \quad To find out the superior algorithms from three or
        more candidate algorithms, the traditional statistical methods
        usually resort to multiple paired \emph{t}-tests. However, it has
        been proved that this approach usually leads to high Type I
        error.\footnote{The probability that we make a mistake to reject the
        null hypothesis, i.e., a misjudgement to say there exists
        significant difference but actually does not.}

        \quad For solving this problem, we turn to the \emph{multiple comparison
        procedure}. The multiple comparison procedure is a statistical test
        technique which helps us compare three or more groups of metrics
        (e.g., classification accuracy) while controlling the probability to make the statistical Type I
        error \cite{pizarro2002multiple}. Moreover, it allows us to concern with a
        set of candidate algorithms not significantly different from the best one
        rather than a single algorithm. Therefore, the multiple comparison procedure is
        an effective method for multi-labeled meta-target collection.

        \quad Therefore, in our experiment, as suggested in \cite{demvar2006statistical}, we employ
        the non-parametric multiple comparison procedure, Friedman followed by Holm's
        procedure test, to obtain the binary-value based meta-target $B_{D}$ for the problem
        $D$ according to $\{Acc_1, Acc_2, \cdots, Acc_{13}\}$ as follows.
        \begin{enumerate}
            \item Applying Friedman test on $\{Acc_1, Acc_2, \cdots, Acc_{13}\}$, the
            null hypothesis of Friedman test is there does not exist
            significant difference among these 13 algorithms. If the result
             of the test support the null hypothesis, all these 13 algorithms
            will be viewed as the appropriate ones. This means that $\forall
            b_i$ of $B_{D}$ ($1\leq i\leq 13$), $b_i = 1$. And the multiple
            comparison procedure is over.

            \item Otherwise, there will exist significant
            difference among these candidate algorithms. In this case, we should apply
            the post-hoc Holm's procedure test to further find the real
            appropriate algorithms. At first, the algorithm with the
            highest average classification accuracy is picked up as a reference.
            And the Holm's procedure test is performed to identify the appropriate
            algorithms from the rest ones. The algorithms that have no significant
            differences with the reference are viewed as the appropriate
            algorithms. Of course, the reference is an appropriate one as
            well. Afterwards, for each value $b_i (1\leq i\leq 13)$ in $B_{D}$, if the corresponding
            algorithm $A_i$ is identified as an appropriate one, $b_i = 1$,
            otherwise, $b_i = 0$.
        \end{enumerate}
    \end{enumerate}
\end{enumerate}

\subsubsection{Recommendation model construction}

The proposed ensemble multi-label learning-based recommendation
model combines a set of base models constructed on different sets of
meta-data. In order to demonstrate whether the proposed ensemble
method is competitive in constructing the recommendation model, we
compare the performance of the ensemble recommendation model with
those of the base models.

When constructing the base model on a given multi-label meta-data,
the data transformation method BR first transforms the multi-labeled
meta-data into multiple single-label meta-data, and then the
well-known classification algorithm, decision tree, is applied on
these single-label meta-data to get the base recommendation
models. The tree-based learner being used is due to the fact that
the it is quite effective to be explored and has good explanation.

Moreover, one critical factor affecting the performance of ensemble
learning is that whether the base models are accurate and diverse.
In order to testify how the accurate and diverse base models affect
the recommendation performance of the ensemble recommendation method
in our experiment, we compare the recommendations of the ensemble
models constructed with respect to four different sets of base
models, including i) all learned base models ii) only accurate base
models, iii) only diverse base models and iv) both of accurate and
diverse models.

\subsubsection{Metrics to Evaluate Recommendation
Models}\label{subsec:evaluateMetrics}

In order to measure the performance of the recommendation model, two
metrics which have been used to evaluate the multi-label methods are
defined as follows.

For a given classification problem $p \in \mathbb{P}$, let $RR_{p} =
<rr_1, rr_2, \cdots, rr_k>$ represent the recommended rank list of
$k$ candidate algorithms $\mathbb{A}$ on $p$. Meanwhile, suppose
that $TB_{p} = <tb_1, tb_2, \cdots, tb_k>$ ($tb_{i} \in \{0, 1\}$)
indicates whether a candidate algorithm $A_i$ is truly appropriate
(i.e., $tb_i = 1$) or not (i.e., $tb_i = 0$) on $p$, $Y$ be the set
of indexes of truly appropriate algorithms and $\hat{Y}$ be the set
of indexes of the inappropriate algorithms on $p$.

Ranking Loss represents the number of times that inappropriate
algorithms are ranked higher than the truly appropriate algorithms.
Ranking loss of the recommended rank list $RR_{p}$ is defined as
follow.

%=================================================================
\begin{definition} Ranking Loss \label{def:rankingLoss}
    \begin{equation}
        R\text{-}Loss(RR_{p}) = \frac{1}{|Y|\cdot |\hat{Y}|}|(i_a, i_b): rr_{i_a} > rr_{i_b},(i_a, i_b) \in Y\times\hat{Y}|
    \end{equation}
\end{definition}
%=================================================================

For algorithm recommendation results in the form of ranking, in a
practical application, the 1st ranked algorithm is usually in favor,
then the 2nd ranked one, and so forth. Therefore, it is natural for
the users to ask that whether the top ranked algorithms are true
appropriate or not. In this case, precision of ranking results,
which has been widely-used in the field of information retrieval to
measure whether the top ranked records are true relevant
\cite{baeza1999modern}, is employed as a measure to evaluate how
well the algorithm-ranking-based recommendation results.

Precision at $m$ to measure the accuracy of the top $m$ recommended
algorithms on problem $D$ is calculated by Eq. \ref{eq:precision}.

%=================================================================
\begin{equation}\label{eq:precision}
        Precision(m) = \frac{number\ of\ real\ appropriate\ algorithms\ within\ top\
        m}{m}.
\end{equation}
%=================================================================

With precision at $m$, average precision \cite{baeza1999modern} to
measure the accuracy of the recommendation result $RR_{D}$ on
problems $D$ is defined as follows.

\begin{definition} Average Precision \label{def:averPrecision}
    \begin{equation}
        AP(RR_{D}) = \sum_{m = 1}^{k}\frac{Precision(m)\times
        \delta(m)}{\sum_{i=1}^{k}tb_i}.
    \end{equation}
\end{definition}

where $k$ denotes the number of the candidate algorithms, and
$\delta(m)$ is a binary function to indicate whether the $m$th
ranked algorithm in $RR_{D}$ is truly appropriate ($\delta(m) = 1$)
or not ($\delta(m) = 0$). The numerator $\sum_{i=1}^{k}tb_i$
represents the number of the truly appropriate algorithms on $D$.

\subsubsection{Recommendation method validation}

After the multi-label meta-data $D_{M}$ with 1090 instances is
acquired, $5\times 10$-fold cross-validation is applied to $D_{M}$ to
empirically evaluate the proposed algorithm recommendation method as follows.

It is important to note that the meta-level evaluation uses held-out
classification problems. During each outer split, the recommendation
model is trained only on the meta-instances in the training fold, the
base-model filtering step uses only the validation fold, and the final
recommendation performance is evaluated only on the test fold. The
meta-targets of the test-fold problems are used only for evaluation and
are not used to train or select the recommendation model.
\begin{enumerate}
    \item $D_{M}$ is randomly divided into 10 sub data sets in the
    same size $\{D_{M_i}: 1\leq i\leq 10\}$, $D_M = \bigcup_{i=1}^{10} D_{M_i}$ and
    $D_{M_i}\cap D_{M_j} = \phi$ ($1\leq i\neq j\leq 10$).
    \item Each sub data set $D_{M_{i}}$ is viewed as the test data
    $D_{te}$, and the union of rest sub data sets $\bigcup_{j=1\wedge j\neq i}^{10} D_{M_j}$
    are randomly divided into two equal-size parts: training data $D_{tr}$ and
    valid data $D_{va}$.
    \item Construct the base recommendation models on the training
    data $D_{tr}$, and filter the base models by their predictions
    on the valid data $D_{va}$ according to Algorithm
    \ref{algo:modelFilter} in Section \ref{subsec:ensembleModel}.
    \item Combine the filtered base recommendation models to form
    the ensemble recommendation model, and evaluate the ensemble
    model in terms of \emph{Ranking Loss} and $\emph{Precision}$ on
    the test data $D_{te}$.
    \item Repeat the above four steps five times, for each time, the
    order of the 1090 instances in $D_{M}$ is rearranged randomly.
\end{enumerate}

\subsection{Results and Analysis}\label{subsec:result}

This section compares the proposed ensemble recommendation model with
individual base recommendation models in terms of \emph{Ranking Loss},
\emph{Average Precision}, and top-ranked recommendation precision. The
analysis emphasizes empirical behavior across the 1,090 classification
problems rather than relying on theoretical guarantees.

For the sake of understanding the results, we denote the different
combinations of the five different types of meta-features in Table
\ref{table:metaFeatureNO}, where numbers ``1'', ``2'', ``3'', ``4''
and ``5'' appearing in column ``Comment'' represent five different
kinds of meta-features, respectively.

%================================================================
\begin{table}[!h]
    \caption{Notations of different combinations of meta-features}\label{table:metaFeatureNO}
    \centering
    \resizebox{0.75\textwidth}{!}{
    \begin{threeparttable}
    \begin{tabular}{l l|l l|l l|l l}
    \hline
    Notation & Comment & Notation & Comment & Notation & Comment & Notation & Comment\\
    \hline
1 & \{1\} & 9 & \{1,5\} & 17 & \{1,2,4\} & 25 & \{3,4,5\}\\
2 & \{2\} & 10 & \{2,3\} & 18 & \{1,2,5\} & 26 & \{1,2,3,4\}\\
3 & \{3\} & 11 & \{2,4\} & 19 & \{1,3,4\} & 27 & \{1,2,3,5\}\\
4 & \{4\} & 12 & \{2,5\} & 20 & \{1,3,5\} & 28 & \{1,2,4,5\}\\
5 & \{5\} & 13 & \{3,4\} & 21 & \{1,4,5\} & 29 & \{1,3,4,5\}\\
6 & \{1,2\} & 14 & \{3,5\} & 22 & \{2,3,4\} & 30 & \{2,3,4,5\}\\
7 & \{1,3\} & 15 & \{4,5\} & 23 & \{2,3,5\} & 31 & \{1,2,3,4,5\}\\
8 & \{1,4\} & 16 & \{1,2,3\} & 24 & \{2,4,5\} &  & \\
    \hline
    \end{tabular}
        \begin{tablenotes}
            \footnotesize
            \item [$\ast$] ``1'' = statistical and information-theory based meta-features;
            ``2'' = model structure based meta-features; ``3'' = Landmarking Based
            meta-features; ``4'' = problem complexity based meta-features and
            ``5'' = structural information based meta-features.
        \end{tablenotes}
    \end{threeparttable}
    }
\end{table}
%===========================================================================

\subsubsection{Comparison on Ranking Loss}

%==================================================================
\begin{figure}[!h]
    \centering
    \includegraphics[width=0.85\textwidth]{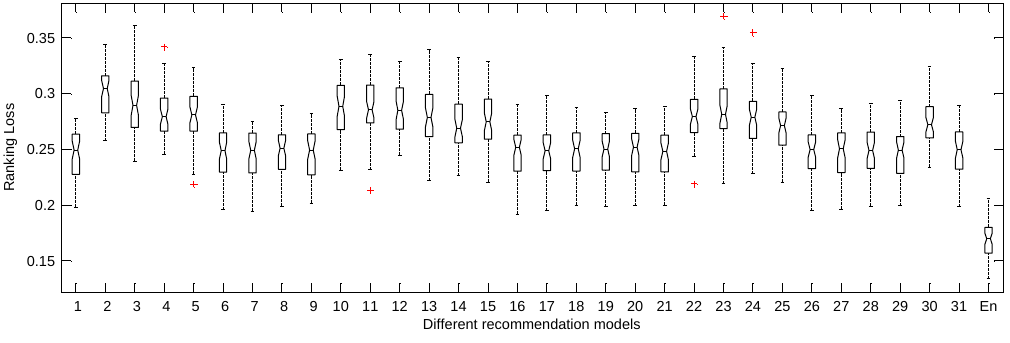}
    \caption{Comparison among different recommendation models in terms of Ranking Loss}\label{Fig:CompOnHammLoss}
\end{figure}
%==================================================================

Fig. \ref{Fig:CompOnHammLoss} compares the ensemble learning-based
recommendation model with the models constructed on different
combinations of meta-features in terms of Ranking Loss. The smaller
the Hamming Loss, the better the corresponding recommendation model.
In this figure, a separate box is produced by the ``box plot'' for
each recommendation model according to its Ranking Loss values
evaluated on the meta-data. The notch of each box denotes the
comparison intervals of the median value of Ranking Loss estimated
on the corresponding recommendation model. Two medians are
significantly different at the 5\% significance level if their
intervals do not overlap. And the box marked as ``En'' denotes the
ensemble learning-based recommendation model, and the $i$th box
denotes of the recommendation model constructed on the $i$th
combination of meta-features in Table \ref{table:metaFeatureNO}. The
same representation can be found in Figs.
\ref{Fig:CompOnMeanPrecision}and \ref{Fig:CompOnPrecision1}. From
Fig. \ref{Fig:CompOnHammLoss}, we can observe that:
\begin{enumerate}
    \item The Ranking Losses of different recommendation models are
    different. And the differences among some models are
    significant. This means that the recommended rankings of candidate algorithms
    vary with different recommendation models. Meanwhile, no matter
    under which kind of combinations of meta-features in Table
    \ref{table:metaFeatureNO}, the recommendation model constructed on the combined meta-features
    performs equally or better than the single kind of
    meta-features. The reason is that different kinds of
    meta-features characterize the classification problems in different
    viewpoints and will be relatively complemented, so the combinations can give us more
    comprehensive understanding of the problem. Furthermore, it is
    possible to construct more precise decision tree model to
    distinguish the appropriate and inappropriate candidate algorithms.

    \item The Ranking Loss of ensemble learning-based model (i.e., the last box) is the
    lowest. And it is significantly smaller than that of any other
    recommendation model (i.e., any box numbered by
    $1,2,\cdots,31$). For the other 31 recommendation models, the
    smallest/greatest median value of Ranking Loss is 0.2483/0.3047.
    However, by combining these 31 recommendation models together
    to form the ensemble recommendation model, the median value of
    Ranking Loss is only 0.17, and outperforms the best base
    recommendation model by 31.53\%. This means that the proposed ensemble learning
    method is more effective to estimate the ranking of the
    candidate algorithms.
\end{enumerate}

\subsubsection{Comparison on Precision}

%================================================================
\begin{figure}[!h]
    \centering
    \includegraphics[width=0.85\textwidth]{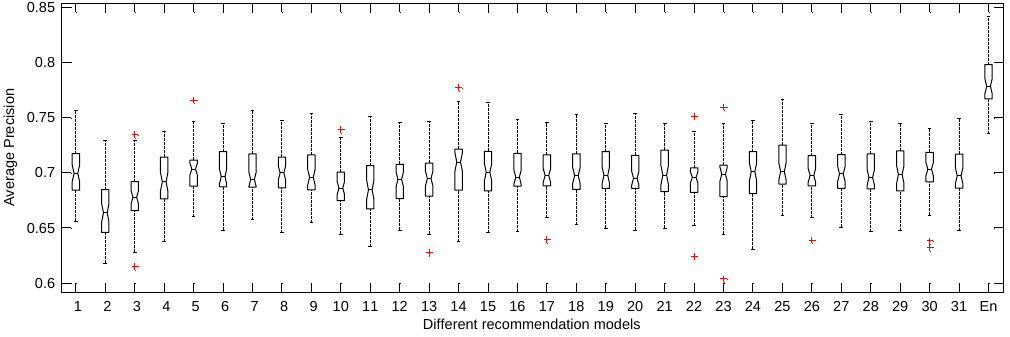}
    \caption{Comparison among different recommendation models in terms of Average Precision}\label{Fig:CompOnMeanPrecision}
\end{figure}
%==================================================================n
Fig. \ref{Fig:CompOnMeanPrecision} shows the comparison results of
different recommendations in terms of average precision. From this
figure, we can get that:
\begin{enumerate}
    \item The Average Precision varies with different recommendation
    models. For single kind of meta-features corresponding to the
    first five recommendation models, there exists distinctly
    significant difference among the Average Precision, such as the
    Average Precision of the models constructed on the statistic and
    information theory and structural information based
    meta-features is significantly greater than that of the models
    constructed on other three kinds of meta-features.

    \quad For the models constructed on the different combinations
    of the five kinds of meta-features, their Average Precisions are
    either statistically equal to or greater than that of model
    constructed on the corresponding single kind of meta-features.
    For example, the 10th box corresponds to the Average Precision
    of the model constructed on the combinations of 2nd and 3rd
    meta-features. And its median is statistically greater than that of either the
    2nd or 3rd box.

    \item The average precision of ensemble learning-based model, which is
    achieved by integrating the 31 recommendation models together, is the
    highest and statistically better than that of any of the other 31
    recommendation models. For the 31 base recommendation models, the
    greatest/smallest median value of Average Precision is 0.7097/0.6637.
    However, by combining these 31 base recommendation models together
    to form the ensemble recommendation model, the median value of
    Average Precision can be up to 0.7785, and outperforms the best base
    recommendation model by 9.69\%. This indicates that combining the
    recommendation models constructed on different sets of
    meta-features together is an effective way to construct the more accurate
    recommendation model.
\end{enumerate}
%==================================================================

%==================================================================
\begin{figure}[!h]
    \centering
    \includegraphics[width=0.85\textwidth]{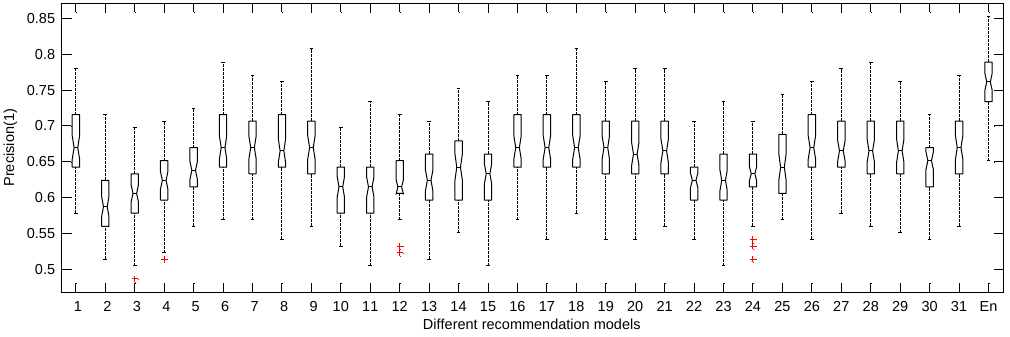}
    \caption{Comparison among different recommendation models in terms of Precision(1)}\label{Fig:CompOnPrecision1}
\end{figure}
%==================================================================

Besides the average precision, the user might be interested on the
precision of the top ranked algorithm. That is, whether the first
recommended algorithm is one of the truly appropriate algorithms.
This can be measured by the metric \emph{Precision(1)} of the
recommendation model. Fig. \ref{Fig:CompOnPrecision1} shows the
\emph{Precision(1)} of different recommendation models. From Fig.
\ref{Fig:CompOnPrecision1}, we can observe that, the Precision of
the top ranked algorithm recommended by different recommendation
models is different. For the 31 base recommendation models, the
greatest/smallest median value of \emph{Precision(1)} is
0.6697/0.5871. However, by combining these 31 base recommendation
models together to form the ensemble recommendation model, the
median value of \emph{Precision(1)} can be up to 0.7615, and
outperforms the best base recommendation model by 13.71\%.

In summary, no matter in terms of either Average Precision or
\emph{Precision(1)}, the proposed ensemble learning-based algorithm
recommendation method is significantly better than the existing
recommendation models.

\subsubsection{Sensitive Analysis of Ensemble Recommendation Model}

%==================================================================
\begin{figure}[!h]
    \centering
    \includegraphics[width=0.85\textwidth]{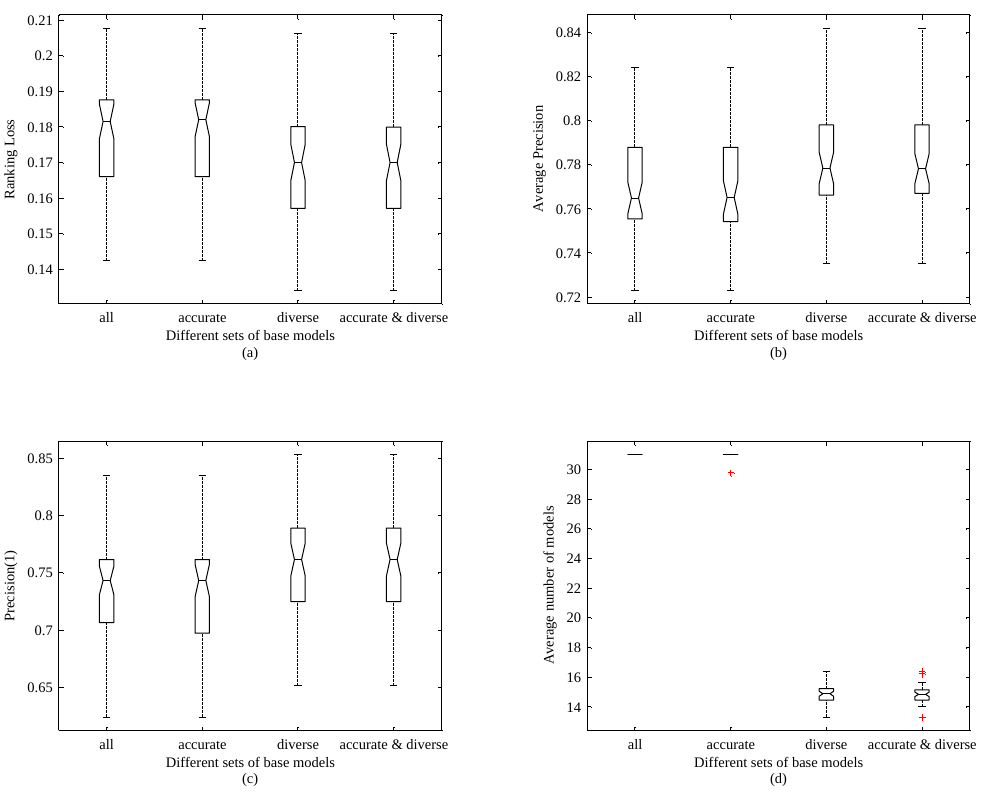}
    \caption{Sensitive Analysis of Accurate and Diverse Models on Ensemble Recommendation Model}\label{Fig:sensitiveAnalysis}
\end{figure}
%==================================================================

A central empirical question is whether accuracy-based and diversity-based
base-model filtering actually improves the ensemble recommendation model.
This section therefore analyzes the contribution of these two factors to
the final recommendation performance. Fig.
\ref{Fig:sensitiveAnalysis} shows the sensitive analyses of these
two aspects on the recommendation models in terms of \emph{Ranking
Loss}, \emph{Average Precision}, \emph{Precision(1)} and \emph{the
number of base models used for ensemble learning},
respectively. In this figure, i) ``all'' represents that all the 31
base recommendation models are used for ensemble model construction;
ii) ``accurate'' denotes the 31 base models are only filtered by the
Definition \ref{def:accurateModel} of accurate learning model; iii)
``diverse'' denotes the 31 base models are only filtered by the
Definition \ref{def:diverseModel} of diverse learning model and iv)
``accurate \& diverse'' means that the 31 base models are filtered
by both Definitions \ref{def:accurateModel} and
\ref{def:diverseModel}.

From Fig. \ref{Fig:sensitiveAnalysis}, we can get that:
\begin{enumerate}
    \item For each sub-figure, the four kinds of recommendation models can be grouped into
    two categories according to the box plots of their
    performance metrics. The models marked as ``all'' and ``accurate''
    have statistically equal performance, and the other two models perform
    statistically equally as well.

    \quad This is because that during the base recommendation model
    construction, for most of the single-label learning problems
    transformed from the multi-label meta-data by BR
    transformation method, the classification accuracy of decision tree on them is greater
    than 0.5. Therefore, the classification accuracy based filter
    can just filter out quite a few decision trees. That mean, for
    recommendation models marked as ``all'' and
    ``accurate'', the decision trees used for model construction are
    quite in common. This leads to that the performance of recommendation
    models marked as ``all'' and ``accurate'' are quite similar.
    Similarly, since the difference among the construction of models
    marked as ``diverse'' and ``accurate \& diverse'' derives from the
    classification accuracy based filter, their performance is
    similar. This can be also confirmed by the Fig.
    \ref{Fig:sensitiveAnalysis} (d) showing the average number of
    trees used for ensemble learning-based algorithm recommendation.
    The recommendation models marked as ``all'' and ``accurate'' are
    constructed based on almost the same number of decision trees.
    So the same as the recommendation models marked as ``diverse'' and ``accurate \&
    diverse''.

    \item Since that the most decision trees used for algorithm
    recommendation model constructed are accurate according to
    Definition \ref{def:accurateModel} of accurate learning model,
    the proposed ensemble learning-based algorithm recommendation
    method focuses on finding out the diverse base learning
    models. According to Fig. \ref{Fig:sensitiveAnalysis}, after
    filtering out the decision trees by Definition
    \ref{def:diverseModel} of diverse base model, we get better
    ensemble learning-based recommendation models with less
    number of decision trees. Such as, the model marked as ``diverse''
    with diversity based filter outperforms the model marked as
    ``all'' in terms of all the performance metrics Ranking Loss,
    Average Precision and Precision(1); and the model marked as
    ``accurate \& diverse'' is better than the model marked as
    ``accurate''. This indicates that $\kappa$ in Eq. \ref{eq:kappa} is a good choice to
    evaluate the diversity between different base models and can
    be used to detect diverse base models for ensemble learning
    based algorithm recommendation.
\end{enumerate}

In summary, by the sensitive analysis of two important aspects
(including accurate and diverse base learning models) in ensemble
learning model construction, we can conclude that, for
classification algorithm recommendation, the proposed definitions of
accurate and diverse learning models are effective to find out a
good set of base recommendation models to construct better ensemble
recommendation model.

\section{Conclusions}\label{sec:conclusion}

This paper proposed a multi-view ensemble meta-learning method for automatically recommending appropriate classification algorithms for new classification problems. The method formulates algorithm recommendation as a multi-label learning problem, reflecting the practical observation that multiple algorithms may be statistically competitive on the same data set.

The main contribution of the paper is empirical. Unlike existing recommendation methods that usually build a single model from one type of meta-feature, the proposed method constructs base recommendation models from different combinations of heterogeneous meta-feature groups and combines them through an accuracy- and diversity-aware ensemble strategy. This design allows the recommendation model to exploit complementary information from multiple views of the data.

We evaluated the proposed method on 1,090 benchmark classification problems, 13 candidate classification algorithms, and five types of meta-features. The experimental results show that the ensemble learning-based recommendation model improves ranking loss, average precision, and top-ranked recommendation precision compared with individual recommendation models. The sensitivity analysis further shows that both accuracy-based and diversity-based base-model filtering contribute to the final recommendation performance.

The proposed framework has several limitations. First, the current implementation uses Binary Relevance to transform the multi-label recommendation problem, which does not explicitly model dependencies among candidate algorithms. Second, the diversity threshold is derived from an approximate statistical criterion and may not be optimal for all data distributions. Third, the empirical study uses classification accuracy as the primary performance measure; other criteria, such as training time, inference cost, robustness, and interpretability, may lead to different recommendations. Finally, the current experiments focus on traditional classification algorithms, and extending the framework to modern AutoML pipelines and deep learning models is an important direction for future work.

Future work will investigate more flexible model-combination strategies, stronger multi-label transformation methods such as classifier chains or RAKEL, additional meta-feature representations, and broader algorithm-selection settings.

\appendix
%\section*{APPENDIX}
\section*{APPENDIX: Meta-features}\label{appendix:meta-features}
\setcounter{section}{1}

Meta-features are measures extracted from a classification problem to describe its properties. These measures map each classification problem into a real-valued vector in the domain $X = R^{m}$ through the extraction function $F$.

Meta-feature extraction is one of the most challenging aspects of algorithm recommendation. In principle, any measure that reflects a property of a classification problem can be viewed as a meta-feature. In practice, however, useful meta-features should be (i) related to the performance of classification algorithms, (ii) easy to compute, and (iii) applicable to different classification problems.

Researchers have proposed several data set characterization methods that describe classification problems from different perspectives. These meta-features can be broadly grouped into the following five categories.

%========================================================================
\begin{table}[!h]
\centering
    \begin{minipage}{0.75\textwidth}
    \small
    \centering
    \renewcommand{\arraystretch}{0.9}
    \caption{Statistical and Information-Theory Based Measures}\label{tab:statMetrics}
    \resizebox{0.75\textwidth}{!}{
        \begin{tabular}{l|l}
        \hline
        Measures & Definitions\\
        \hline
        Ins.Num & Number of instances\\
        Attr.Num & Number of Attributes\\
        Target.Num & Number of target concept values\\
        Target.Min & Proportion of minority target\\
        Target.Max & Proportion of majority target\\
        Pro.Bin & Proportion of binary attributes\\
        Pro.Nom & Proportion of nominal attributes\\
        Pro.Num & Proportion of numeric attributes\\
        Pro.MissIns & Proportion of instances with missing values\\
        Pro.MissValues & Proportion of missing values\\
        \hline
        Mean.Geo & Geometric mean\\
        Mean.Harm & Harmonic mean\\
        Mean.Trim & Trim mean excluding the highest and lowest 5\%\\
        Mad & Mean absolute deviation\\
        Var & Variance\\
        Std & Standard deviation\\
        Prcitile & Percentile 75\%\\
        Int.Range & Interquartile range\\
        Prop.AttrWithOutlier & Proportion of numerical attributes with outliers over all numerical attributes\\
        Skewness & Skewness of data based on numerical attributes\\
        Kurtosis & Kurtosis of data based on numerical attributes \\
        Max.eig & Maximum eigenvalue\\
        Min.eig & Minimum eigenvalue\\
        Can.corr & Canonical correlation\\
        Grav.cent & Center of gravity\\
        MeanAbsCoef & Mean absolute coefficient of attribute pairs \\
        \hline
        \emph{$H(C)$} & Entropy of classes \\
        \emph{$\bar{H}(X)$} & Mean entropy of nominal attributes \\
        \emph{$\bar{M}(C,X)$} & Mean mutual information of classes and attributes based on nominal attributes\\
        En.attr &  Equivalent number of attributes \emph{$H(C)/\bar{M}(C,X)$}\\
        Ns.ratio & Noise-signal ratio \emph{$\bar{H}(X)/\bar{M}(C,X)-1$} \\
        \hline
        \end{tabular}
    }
    \end{minipage}
    \begin{minipage}{0.75\textwidth}
    \footnotesize
    \centering
    \renewcommand{\arraystretch}{0.9}
    \caption{Model Structure Based Measures}\label{tab:modelMetrics}
    \resizebox{0.6\textwidth}{!}{
        \begin{tabular}{l|l}
        \hline
        Measures & Definitions\\
        \hline
        Tree.Height & Height of tree (also referred as to number of levels in tree)\\
        Tree.Wdith & Width of tree\\
        Node.Num & Number of nodes in tree\\
        Leaf.Num & Number of leaves in tree\\
        \hline
        Level.Max & Maximum number of nodes at one level\\
        Level.Mean & Mean of the number of nodes on levels\\
        Level.Dev & Standard deviation of the number of nodes on levels\\
        \hline
        Branch.Long & Length of the longest branch\\
        Branch.Short & Length of the shortest branch\\
        Branch.Mean & Mean of the branch lengths\\
        Branch.Dev & Standard deviation of the branch lengths\\
        \hline
        Attr.Min & Minimum occurrence of attributes\\
        Attr.Max & Maximum occurrence of attributes\\
        Attr.Mean & Mean of the number of occurrences of attributes\\
        Attr.Dev & Standard deviation of the number of occurrences of attributes\\
        \hline
        \end{tabular}
    }
    \end{minipage}
    \begin{minipage}{0.75\textwidth}
    \footnotesize
    \centering
    \renewcommand{\arraystretch}{0.9}
    \caption{Problem Complexity Based Measures}\label{tab:complexityMetrics}
    \resizebox{0.6\textwidth}{!}{
        \begin{tabular}{l|l}
        \hline
        Measures & Definitions\\
        \hline
        Bound.Len & Length of class boundary \\
        Adherence.Prop &  Proportion of retained adherence subsets\\
        Intra/Inter.Ratio &  Ratio of average intra/interclass nearest neighbors \\
        NN.Nonlinerity &  Nonlinearity of Nearest Neighbors classifier \\
        Linear.Nonlinerity & Nonlinearity of linear classifier \\
        Fisher.Ratio & Maximum Fisher's discriminant ratio \\
        Ins/Attr & Training set size relative to feature space dimensionality \\
        \hline
        \end{tabular}
    }
    \end{minipage}
\end{table}
%========================================================================

\begin{enumerate}
    \item \emph{Statistical and Information-theory Based Measures}

    The statistical and information-theory based measures are the most
    widely-used in the field of classification algorithm recommendation
    \cite{brazdil2003ranking,sohn1999meta,Henery1995methods,aha1992generalizing,gama2000cascade,Engels98usinga}. The prominent
    examples based on these measures are the projects ESPRIT Statlog
    (1991-1994) and METAL (1998-2001). These measures generally include
    the data set characteristics such as, number of features, number of
    instances, number of target concepts, ratio of instances to
    features, ratio of missing values, ratio of binary features, entropy
    of the target concept, information gain between the feature and the
    target concept, and correlation coefficient between features,
    etc. See Table \ref{tab:statMetrics} for details.

    \item \emph{Model Structure Based Measures}

    Firstly, a classification problem is represented in a
    special data structure embedding the complexity of the problem.
    Then, the characteristics of the structure are exploited to describe
    the classification problem.

    \quad In the field of algorithm recommendation, the induced decision
    tree is a well-known and commonly-used structure to model a
    classification problem. Bensusan \cite{Bensusan1998god} proposed to capture the
    information from the induced decision tree for describing the
    classification complexity. He extracted ten measures from the decision
    tree, such as the ratio of the number of nodes to the number of
    features, the ratio of the number of nodes to the number of
    instances, etc. Afterwards, Peng et al. \cite{peng2002improved}
    re-analyzed the characterization of decision trees, and proposed
    some new measures to characterize the structural properties of
    decision trees. See Table \ref{tab:modelMetrics} for details.

    \item \emph{Landmarking Based Measures}

    This kind of measures falls within the concept of landmarking
    \cite{Pfahringer00meta,Bensusan2000casa,jain2000statistical,duin2004characterization}.
    This idea was proposed based on the assumption that the
    performance of the candidate algorithms could be predicted by
    the performance of a set of simple classifiers (also called
    landmarkers). So the performance (e.g., accuracy) of these landmarkers
    is used to describe a classification problem. Evidently, this kind of measures is closely related to
    the choice of landmarkers. In practice, it should be ensured that the
    chosen landmarkers have significant differences in terms of learning
    mechanism. Following the suggestions in
    \cite{Bensusan2000casa,Pfahringer00meta}, the following six classifiers are
    selected as the landmark learners: i) Naive Bayes, ii) 1-NN (Nearest Neighbor),
    iii) Elite 1-NN, iv) a decision node tree, v)
    a random chosen node tree and  vi) the worst node tree.
    Where the last three classifiers can be achieved based on the
    well-known classification algorithm C4.5.

    \item \emph{Problem Complexity Based Measures}

    The problem complexity based measures focus on the
    description of the geometrical complexity of the classification problem and
    emphasize the geometrical characteristics of the distributions of
    the classes by analyzing the source of difficulty in solving a classification problem
    \cite{bernado2005domain,ho2002complexity,elizondo2009estimation,ho2000complexity}.
    The measures reflecting the way in which different
    classes are separated or interleaved (and being relevant to
    classification performance) are identified as the measurement of the problem's
    complexity. Such as Fisher's discriminant ratio, the percentage of
    instances in the problem that linear the class boundary, and the
    nonlinearity of linear/non-linear classification algorithm, etc. See Table
    \ref{tab:complexityMetrics} for details.

    \item \emph{Structural Information Based Measures}

    Recently, Song et al. proposed a novel data characterization
    method to facilitate the algorithm recommendation
    \cite{song2012automatic}. The method utilizes structural information
    based feature vectors to characterize the classification problems, which
    is quite different from the existing ones. Specially, the two
    feature vectors, one-item feature vector and two-item feature
    vector, are extracted from a given classification problem. These two vectors
    consist of the frequencies of one-item sets and two-item sets,
    respectively. Afterwards, the \emph{minimum}, \emph{1/8
    quantile}, \emph{2/8 quantile}, \emph{3/8 quantile}, \emph{4/8
    quantile}, \emph{5/8 quantile}, \emph{6/8 quantile}, \emph{7/8
    quantile} and \emph{maximum} are computed for these two vectors
    and form the final set of data set characteristics.
\end{enumerate}

%\appendixhead{ZHOU}
%
%% Acknowledgments
%\begin{acks}
%The authors would like to thank Dr. Maura Turolla of Telecom
%Italia for providing specifications about the application scenario.
%\end{acks}

% Bibliography
\bibliographystyle{acmsmall}
\bibliography{AlgRecOnEnsemble}
                             % Sample .bib file with references that match those in
                             % the 'Specifications Document (V1.5)' as well containing
                             % 'legacy' bibs and bibs with 'alternate codings'.
                             % Gerry Murray - March 2012

% History dates
%\received{February 2007}{March 2009}{June 2009}
%
%% Electronic Appendix
%\elecappendix

\medskip

%\section{This is an example of Appendix section head}
%
%Channel-switching time is measured as the time length it takes for
%motes to successfully switch from one channel to another. This
%parameter impacts the maximum network throughput, because motes
%cannot receive or send any packet during this period of time, and it
%also affects the efficiency of toggle snooping in MMSN, where motes
%need to sense through channels rapidly.
%
%By repeating experiments 100 times, we get the average
%channel-switching time of Micaz motes: 24.3 $\mu$s. We then conduct
%the same experiments with different Micaz motes, as well as
%experiments with the transmitter switching from Channel 11 to other
%channels. In both scenarios, the channel-switching time does not have
%obvious changes. (In our experiments, all values are in the range of
%23.6 $\mu$s to 24.9 $\mu$s.)
%
%\section{Appendix section head}
%
%The primary consumer of energy in WSNs is idle listening. The key to
%reduce idle listening is executing low duty-cycle on nodes. Two
%primary approaches are considered in controlling duty-cycles in the
%MAC layer.

\end{document}